\documentclass[twocolumn,numberedappendix]{emulateapj}
\usepackage{graphicx}
\usepackage{hyperref}
\usepackage{amssymb}
\usepackage{amsmath}
\usepackage{units}
\newcommand{\figwidth}{3.35in}

\newcommand{\twopartdef}[4]
{
  \left\{
    \begin{array}{ll}
      #1 & \mbox{if } #2 \\
      #3 & \mbox{if } #4
    \end{array}
  \right.
}

\newcommand{\comments}[1]{}

\begin{document}

\title{Resonance Broadening and Heating of Charged Particles in
  Magnetohydrodynamic Turbulence}
\author{Jacob W. Lynn\altaffilmark{1,2}, Ian J. Parrish\altaffilmark{2},
  Eliot Quataert\altaffilmark{2}, \& Benjamin D. G. 
  Chandran\altaffilmark{3}} 

\altaffiltext{1}{Physics Department, University of California,
  Berkeley, CA 94720; jacob.lynn@berkeley.edu} 
\altaffiltext{2}{Astronomy Department and Theoretical Astrophysics 
  Center, University of California, Berkeley, CA 94720} 
\altaffiltext{3}{Space Science Center and Department of Physics,
  University of New Hampshire, Durham, NH 03824}

\begin{abstract}

  The heating, acceleration, and pitch-angle scattering of charged
  particles by MHD turbulence are important in a wide range of
  astrophysical environments, including the solar wind, accreting
  black holes, and galaxy clusters. We simulate the interaction of
  high-gyrofrequency test particles with fully dynamical simulations
  of subsonic MHD turbulence, focusing on the parameter regime with
  $\beta \sim 1$, where $\beta$ is the ratio of gas to magnetic
  pressure. We use the simulation results to calibrate analytical
  expressions for test particle velocity-space diffusion coefficients
  and provide simple fits that can be used in other work. 

  The test particle velocity diffusion in our simulations is due to a
  combination of two processes: interactions between particles and
  magnetic compressions in the turbulence (as in linear transit-time
  damping; TTD) and what we refer to as Fermi Type-B (FTB) interactions,
  in which charged particles moving on field lines may be thought of as
  beads sliding along moving wires. We show that test particle heating
  rates are consistent with a TTD resonance which is broadened
  according to a decorrelation prescription that is Gaussian in time
  (but inconsistent with Lorentzian broadening due to an exponential
  decorrelation function, a prescription widely used in the
  literature). TTD dominates the heating for $v_s \gg v_A$
  (e.g. electrons), where $v_s$ is the thermal speed of species s and
  $v_A$ is the Alfv\'{e}n speed, while FTB dominates for $v_s \ll v_A$
  (e.g. minor ions). Proton heating rates for $\beta \sim 1$ are
  comparable to the turbulent cascade rate. Finally, we show that
  velocity diffusion of collisionless, large gyrofrequency particles
  due to large-scale MHD turbulence does not produce a power-law
  distribution function. 
\end{abstract}


\section{Introduction}
\label{sec:introduction}

The interaction between charged particles and magnetohydrodynamic
(MHD) turbulence plays a role in the energy balance of diverse
astrophysical environments such as the solar corona and solar wind 
\citep[e.g][]{Cranmer2007} and accretion disks around black holes
\citep[e.g.][]{Quataert1999}. The coupling between turbulence and
particles is also important for the transport and confinement of
cosmic rays in galaxies \citep[e.g.][]{Chandran2000,Yan2004}.

This paper focuses on the interaction of test particles with subsonic
(and thus weakly compressible) MHD turbulence. Such turbulence
consists primarily of nonlinearly interacting Alfv\'{e}n 
waves which drive the turbulent cascade
\citep{Kraichnan1965,Shebalin1983,Goldreich1995}, along 
with slow magnetosonic modes that are advected passively by the
Alfv\'{e}nic cascade \citep{Lithwick2001,Cho2002}. On 
observational \citep{Chen2011b}, theoretical \citep{Goldreich1995},
and numerical \citep{Maron2001,Beresnyak2011c}
grounds, the MHD cascade is believed to 
be strongly anisotropic, with most power in the inertial range of the
cascade in modes with wavevectors primarily perpendicular to the
magnetic field. This has many
implications for the coupling between particles and turbulence,
including e.g. that cyclotron heating
of particles by the MHD cascade is significantly suppressed
\citep[][though see \cite{Leamon1999,Gary2008,Jiang2009} for
discussions of the cyclotron resonance in the corona and solar
wind]{Quataert1998,Cranmer2003}.  

The interaction between test particles and plasma waves has been 
extensively studied in the ``quasilinear'' approximation
\citep{Kennel1966, Jokipii1966}, which 
stipulates that test particles execute unperturbed helical motion
around magnetic field lines, and that plasma waves are
long-lived relative to their periods. This
implies that wave-particle interactions and energy exchange occur only
at discrete 
resonances \citep[and references therein]{Stix1992}. To the extent
that MHD turbulence is well-described by a superposition of
long-lived small-amplitude plasma 
waves, quasilinear theory will accurately describe test particle
diffusion and heating in turbulence.

However, the picture of strong anisotropic MHD turbulence developed
over the last $\sim \! \! 30$ years by many authors 
\citep[e.g.][henceforth GS]{Montgomery1981,Shebalin1983, Higdon1984, Goldreich1995} suggests
that the Alfv\'{e}n and slow waves comprising weakly compressible MHD
turbulence are not long-lived, 
and instead decorrelate due to non-linear interactions before they can
propagate over distances of 
multiple wavelengths.\footnote{The use of \textit{strong} here refers to this
  state of ``critical balance'' 
  between eddy and wave timescales, rather than the amplitude of the
  turbulence.} As a result, the discrete resonances of quasilinear
theory are expected to be substantially broadened in MHD turbulence
\citep[e.g.][]{Bieber1994a,Shalchi2004,Shalchi2004a,Qin2006,Yan2008a}.

In this paper, we study the interaction between test particles and
driven MHD turbulence \citep[see also earlier work by][]{Dmitruk2004,
  Lehe2009}. In particular, we
quantify the velocity-space  diffusion, particle heating, and particle
acceleration that results.   We  compare these numerical  results in
detail to analytic models and, in  particular, calibrate models of
resonance broadening.   Our focus is on  the basic physics---the
diffusion coefficients we calculate
(e.g. eqn.\ \ref{eq:exponentialDecorrelation} \&
\ref{eq:GaussianDecorrelation} \& Table \ref{table:Ci}) can
be used for a wide range of 
applications, some of which  we will  explore in detail in 
future studies. 

The paper is structured as follows. In \S
\ref{sec:qualitativeTransport}, we summarize the qualitative features
of test particle interactions with turbulence, including both
resonance broadening and non-resonant interactions. In \S
\ref{sec:transportProperties} we use these physical ideas to derive
analytical expressions for velocity diffusion coefficients in 
turbulence, while in \S \ref{sec:stochasticHeating} we calculate the
resulting heating rates for 
a thermal distribution of test particles. Many of these results are
not new, but they provide a useful analytic framework for interpreting
our test particle numerical results and so are included for
completeness. In \S
\ref{sec:numericalMethods} we describe our numerical methods for
evolving test particles in simulations of MHD turbulence, and in \S
\ref{sec:particleTransport} we  compare our analytical predictions
to the results of test particle simulations. Finally, in \S
\ref{sec:conclusions} we discuss the conclusions and implications of
our work. We also include several Appendices which consider related
ideas. In Appendix \ref{app:oneWave}, we discuss the interaction of
test particles with one finite-amplitude wave, in Appendix
\ref{app:weakStrong}, we consider an extension of our model into a
the regime of weak turbulence, and in Appendix \ref{app:powerSpectra},
we discuss the power spectra of our turbulence simulations in the
context of weak and strong turbulence.

\section{Qualitative Discussion of Test Particle Transport in MHD
  Turbulence}
\label{sec:qualitativeTransport}

In this paper, we focus on isothermal MHD
turbulence with $\beta = \rho c_s^2 / (B_0^2 / 8 \pi) \sim 1$ and
$\dot{\epsilon}$ such that the 
turbulence is subsonic and sub-Alfv\'{e}nic, where
$\dot{\epsilon}$ is the energy input rate per unit mass into the 
turbulence. Additionally, we focus on high-gyrofrequency particles
with $\Omega \gg \omega_{\mathrm{max}}$, where  
$\Omega = q B / m c$ is the particle cyclotron frequency and
$\omega_{\mathrm{max}}$ is the 
maximum resolveable wave frequency in the turbulence. Our motivation
for doing so is that this  inequality is believed to be satisfied even
deep in the inertial range of weakly  compressible MHD turbulence
(see e.g. \citealt{Howes2008}).
In addition to
$\Omega \gg \omega_{\mathrm{max}}$, magnetic moment conservation
generally also 
requires that the amplitude of  the turbulent fluctuations on scales
of the Larmor radius $r_L$ satisfy  $\delta v/v_\perp \ll 1$
\citep{McChesney1987,Chaston2004,Chandran2010}.   This is  satisfied
in our simulations 
both because we focus on subsonic $\beta \sim 1$ turbulence and
because the 
Kolmogorov power spectrum  that the turbulence self-consistently
develops has only modest power on  small scales $\sim r_L$.   We shall
see that magnetic moment  conservation is indeed reasonably well
satisfied in our 
test particle simulations (see \S \ref{sec:particleTransport}). This
implies that particle acceleration will be primarily in the parallel
direction.  (We use the subscripts $\parallel$ and $\perp$
throughout to indicate parallel and perpendicular to the local
magnetic field, respectively.) 

Generically, there will be two processes that cause changes in
parallel velocity for the high-$\Omega$ particles under
consideration. The first is 
\textit{transit-time damping} (TTD), 
which is analogous to Landau damping. In a spatially-varying
magnetic field, charged particles feel mirror forces, given by $\mu
\nabla_\parallel B$, where $\mu = m v_\perp^2 / 2 B$ is the particle's
magnetic moment. If the spatial variation is provided by a
compressive wave moving with a phase speed $v_p$, then particles
with $v_\parallel \simeq v_p$ will ``surf'' the wave and experience
correlated acceleration for long times (until the particle is
accelerated such that it is no longer in resonance). On the other
hand, a particle with $v_\parallel \not \simeq v_p$ will experience
time-varying accelerations that will average to zero over long
times. Thus, over time scales sufficiently short that particle  
velocities do not change substantially (where linear theory applies)
the interaction of the wave with a distribution 
of particles will be given by a delta-function, $D_\parallel \propto
\delta(v_\parallel - 
v_p)$, where $D_\parallel$ is the parallel velocity diffusion coefficient.

This picture will be modified for the interaction of particles with
strong turbulence. Strong turbulence can be thought of as a
distribution of waves, which nonetheless do not propagate long
distances as waves, but instead decohere on a timescale 
$\omega^{-1}$, where $\omega$ is the frequency of the wave. The linear
theory model for the interaction  of particles and waves is only valid
for waves  
that are relatively long-lived; qualitatively, wave decoherence will
cause the delta-function resonance to broaden, with more particles
able to approximately satisfy the resonance condition for long enough
to experience significant acceleration.

\begin{figure}
  \begin{center}
    \includegraphics[width=\figwidth]{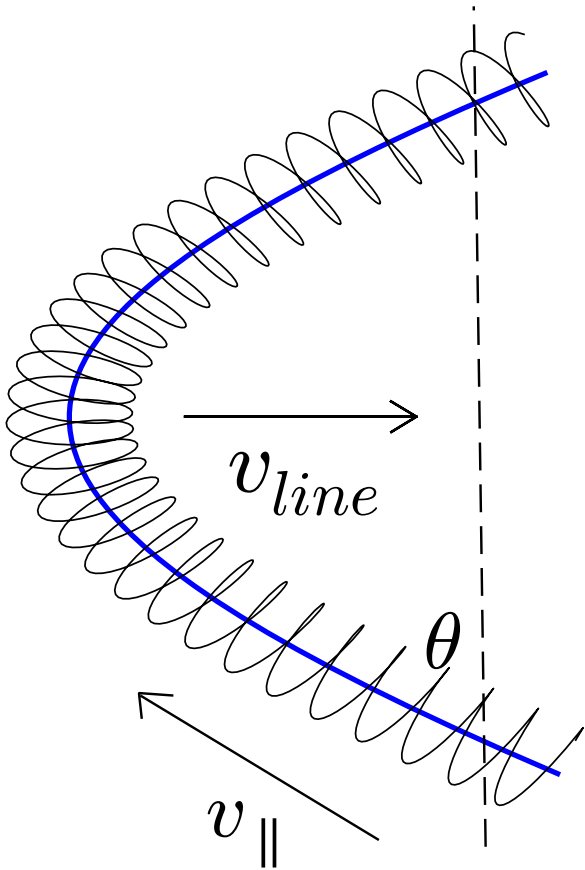}
    \caption{In a Fermi Type-B interaction, a particle moving with
      parallel velocity $v_\parallel$ 
      along a curved element of magnetic field line will gain or lose
      parallel velocity $\Delta v_\parallel = v_{\mathrm{line}}
      \sin{\theta}$, assuming the pictured geometry, due to being slung
      around like a ``bead on a wire.''}
    \label{fig:FTB}
  \end{center}
\end{figure}

Because TTD arises from the mirror force, it will
become negligible as $\mu \rightarrow 0$. In this limit (and in the
high-$v_\parallel$ limit, as we will show) the most important
mechanism for changes in parallel velocity is what we refer to as
\textit{Fermi Type-B} (FTB) acceleration
\citep{Fermi1949a}. Consider a particle 
spiralling along a magnetic field line with some curvature (see Fig.\
\ref{fig:FTB}). In the 
frame of the field line, the particle has constant energy, because
magnetic fields do no work. In the frame of the bulk plasma (i.e., the
frame in which the average  
fluid momentum is zero), however, the particle may gain or
lose energy. This can be seen straightforwardly with a Galilean
transform (in the non-relativistic case) from the field line frame to
the plasma frame. Qualitatively, FTB describes charged particles
as beads sliding along moving wires. These stochastic
interactions will also cause diffusion in velocity space independent 
of $\mu$.

As noted above, we do not consider particles with $\Omega \sim
\omega$. Particles with $\omega - k_\parallel v_\parallel = n
\Omega$, where $n$ is a non-zero integer, are
``cyclotron-resonant'' and can thus experience violation of
$\mu$-conservation 
and strong perpendicular heating. However, due to the anisotropy of
the strong MHD turbulent cascade \citep{Goldreich1995}, MHD turbulence 
with a substantial inertial range may transfer its energy 
into a kinetic Alfv\'{e}n wave cascade on scales of
$k_{\perp,\mathrm{max}} \sim r_L^{-1}$, where $r_L$ is the proton
gyroradius. At this scale in the solar wind, the maximum gyrofrequency
that can be  cyclotron-resonant for thermal particles is $\ll
\Omega_p$, the proton gyrofrequency (see  \citealt{Lehe2009}). As a
result, any cyclotron heating of protons or electrons 
occurs (if it occurs at all) on scales below the MHD cascade that we
consider here. However, there is recent work
\citep{Cranmer2012} suggesting that mode coupling between the
isotropic fast mode cascade and Alfv\'{e}n modes 
could supply enough high-$k_\parallel$ power in the solar wind to heat
protons through the cyclotron resonance.

Our test particle calculations presented in \S
\ref{sec:transportProperties} do not include 
diffusion due to parallel electric fields (Landau damping).   The
reason is that our simulations are performed in the ideal MHD limit.
Physically, sound waves in a collisionless plasma do generate
significant parallel electric fields -- this effect is potentially
important for the fast mode in $\beta \gtrsim 1$ MHD turbulence and
for the slow mode in $\beta \lesssim 1$ turbulence, but is not
captured in our test particle calculations. Additionally, the ideal
MHD 
limit does not allow us to capture heating and acceleration by
magnetic reconnection, which is likely an important mechanism in
many astrophysical environments including solar flares \citep[see
e.g.][]{Drake2006,Drake2009}. 

\section{Analytic Transport Properties}
\label{sec:transportProperties}

In this section, we provide an analytic derivation of the 
evolution of a distribution of test particles in velocity space as
they interact with Alfv\'{e}nic MHD turbulence. We focus on the diffusive 
evolution in $v_\parallel$, which dominates over any changes in
$v_\perp$, as argued in \S \ref{sec:qualitativeTransport}.

\subsection{Transport in $v_\perp$}
\label{sec:vperpTransport}

The magnetic moment $\mu = v_\perp^2/B$ is approximately a conserved
quantity in our test particle calculations described in \S
\ref{sec:numericalMethods} (for more discussion, see \S
\ref{sec:qualitativeTransport}).

If $\mu$ is conserved, then any change in $v_\perp$ results
from a
change in the local value of the magnetic field. A distribution of
particles that initially all have the same value of $v_\perp$,
randomly initialized in space throughout a turbulent plasma, will
quickly broaden in $v_\perp$ over short time scales, and
then reach a saturated width in $v_\perp$ as 
the particles statistically sample all of the fluctuations in
$B$. Taking a differential of $v_\perp = \sqrt{\mu B}$, we find that 
final width of the distribution in $v_\perp$ will 
be approximately given by $\delta v_\perp / v_\perp \sim B_L /
2 B_0$, where $B_L$ is the rms deviation of $B$ from the mean. We can
understand this more simply by noting that if a particle is
initialized where $B$ is higher than $B_0$, then $\mu$ for that
particle will be lower than the average. Thus the fractional width of
the initial distribution in 
$\mu$ will be the same as the width of the initial distribution in
$B$ (in the $B_L/B_0 \ll 1$ limit). After some time passes and the
test particles statistically 
sample the turbulence, a particle with a smaller $\mu$ is not
preferentially likely to be in a location of larger magnetic field, so
this particle will have a smaller time-averaged $v_\perp$. To
the extent that $\mu$ is well-conserved, there will be little 
diffusion in $v_\perp$ after this initial re-adjustment.

\subsection{Transport in $v_\parallel$}

Diffusion in parallel velocity in our simulations comes from two
sources, which \cite{Fermi1949a} referred to as ``Type A'' and ``Type
B'' interactions. Type A refers to acceleration by magnetic mirror 
forces. In Fermi's original conception, charged particles 
reflected off of magnetic inhomogeneities (clouds) due to the $\mu 
\nabla B$ force. In this work, the scattering
centers are compressive MHD modes, so that Type A acceleration
and TTD are effectively the same. Type B refers to the
acceleration of a particle in the rest frame of the bulk plasma, due
to following a curved, moving field line.

\subsubsection{TTD and Resonance Broadening}
\label{sec:broadening}

A particle moving in compressible MHD turbulence will be randomly
accelerated and decelerated by mirror forces, given by $\mu \nabla
B$. This will lead to diffusion in parallel velocity given by
\begin{equation}
  \label{eq:fokkerPlanck}
  \frac{\partial f}{\partial t} = \frac{\partial}{\partial
    v_\parallel} \left( D_\parallel \frac{\partial f}{\partial
      v_\parallel} \right),
\end{equation}
where $f(\mathbf{x},\mathbf{v})$ is the velocity-space distribution
function (we will assume a uniform spatial distribution of test
particles throughout). We adopt the formal approach of \cite{Dupree1966} and
\cite{Weinstock1969} and write the corresponding diffusion coefficient
as 
\begin{equation}
  \label{eq:DAvg}
  D_\parallel = \mu^2 \int_0^\infty dt ~ \langle \nabla_\parallel B
  (\mathbf{r}_0, 0) \nabla_\parallel B (\mathbf{r(t)}, t) \rangle, 
\end{equation}
where $\langle \rangle$ indicates an ensemble average,
$\nabla_\parallel$ means the gradient in the parallel (to the magnetic
field) direction, and
$\mathbf{r}(t)$ describes a particle's trajectory. Introducing a
Fourier transform of the magnetic field, we can then write
Equation \ref{eq:DAvg} as
\begin{equation}
  \label{eq:DSpectrum}
  D_\parallel = \frac{\mu^2}{(2 \pi)^3} \int
  d^3 k ~
  k_\parallel^2 I(\mathbf{k}) R(\mathbf{k}),
\end{equation}
where $I(\mathbf{k})$ is the power spectrum of magnetic field
fluctuations, and $R(\mathbf{k})$ is a resonance function. If we
consider only free-streaming trajectories (ignoring any variations in
$v_\parallel$), as in linear theory, the resonance function is given by
\begin{equation}
  \label{eq:pureResonance}
  R(\mathbf{k}) = \operatorname{Re}{\int_0^\infty dt
    ~e^{\imath(\omega(\mathbf{k}) - v_\parallel k_\parallel) t}}.
\end{equation}
This resonance function becomes the standard $\delta (v_\parallel \pm
v_p)$ for interactions with waves satisfying the linear
dispersion relation $\omega(\mathbf{k}) =
\pm |k_\parallel| v_p$. The $\mu
\nabla_\parallel B$ force 
requires compressive fluctuations, and thus will stem  from slow
or fast waves 
in MHD turbulence, rather than Alfv\'{e}n waves. However,
the slow mode cascade passively adopts the anisotropic GS power
spectrum, with $k_\perp \gg k_\parallel$ in the inertial range
\citep{Cho2002}. In this regime, the slow mode dispersion relation
becomes $\omega = k_\parallel c_s v_A/ \sqrt{c_s^2+v_A^2}$
\citep{Chandran2003a},
limiting to $\omega = k_\parallel v_A$ for $\beta \gg 1$ and $\omega =
k_\parallel c_s$ for $\beta \ll 1$. For analytical simplicity, we 
use a generic Alfv\'{e}nic dispersion relation, $\omega =
k_\parallel v_p$, where $v_p$ is the appropriate phase velocity. The
specification of the results presented here to various slow mode
regimes is straightforward.

Deviation 
from free-streaming trajectories modifies the resonance function in
Equation \ref{eq:pureResonance} \citep{Weinstock1969}; the resonance function becomes
\begin{equation}
  \label{eq:weinstockResonance}
  R(\mathbf{k}) = \operatorname{Re}{\int_0^\infty dt
    ~e^{\imath(\omega(\mathbf{k}) - v_\parallel k_\parallel) t +
      \frac{1}{2}\langle [\imath \mathbf{k} \cdot \delta \mathbf{r}(t)]^2\rangle}},
\end{equation}
where $\delta \mathbf{r}(t)$ describes the deviation of a particle's
trajactory, and we have assumed the mean deviation $\langle
\delta\mathbf{r}(t) \rangle = 0$. Formally, this 
result requires $k \tau_B^2 \mu \nabla_\parallel B \ll 1$, where
$\tau_B$ is the correlation time of the stochastic 
magnetic field fluctuations on a given scale $k$, which will not hold
for all scales or for all test particles we
consider. Furthermore, the exact value of $\langle
|\delta\mathbf{r}(t)|^2 
\rangle$ depends on the diffusion coefficient that we are trying to
calculate, and would require a recursive approach to calculating
$D_\parallel$ and $D_\perp$. Thus, for analytical simplicity, we use
Equation \ref{eq:weinstockResonance} only to motivate a phenomenological
modification to the 
resonance function.

To estimate an appropriate modification, consider that one source
of deviation from free-streaming trajectories is the fact 
that high-$\mu$ particles are tied to magnetic field lines that can
wander in the perpendicular direction. This process will be
essentially stochastic and therefore diffusive. The typical step
length will be given by $v_l \tau_B$, where
$v_l$ is the scale-dependent rms turbulent velocity. Thus:
\begin{equation}
  \label{eq:FLRWEstimate}
  \langle \delta \mathbf{r}_\perp^2 (t) \rangle \sim
  (v_l \tau_B)^2 \left(\frac{t}{\tau_B}\right) \sim
  v_l^2 \tau_B t.
\end{equation}

Throughout the main body of the paper we focus on the strong
anisotropic turbulence model of  
\citet[henceforth GS]{Goldreich1995}, though we consider an extension
to weak turbulence in Appendix \ref{app:weakStrong}.\footnote{Note
  that since we 
  drive our test particle simulations in \S
  \ref{sec:particleTransport} at 
  sub-Alfv\'{e}nic velocities on the outer scale, we are in fact in the
  weak/intermediate 
  turbulence regime on size
  scales of $l_\perp \gtrsim l_c \equiv L M_A^2$, where $L$ is the outer
  (driving) scale of the turbulence and $M_A$ refers to the Alfv\'{e}nic
  Mach number at $L$ \citep[see
  e.g.][]{Goldreich1997,Galtier2000a}. We discuss an extension of the
  model in this section which incorporates weak turbulence in Appendix
  \ref{app:weakStrong}.} In the 
GS model, the correlation time of 
fluctuations is given by $\tau_B^{-1} \sim \omega_{\mathrm{nl}} =
(k_\perp L)^{2/3} v_L/L$, where $v_L$ is the rms turbulent velocity on
the driving scale $L$. This timescale may be thought of as the
``lifetime'' of waves in turbulence. Thus our estimate for the field
line random-walk deviation becomes  
\begin{equation}
  \label{eq:FLRWEstimate2}
  \langle [\mathbf{k} \cdot \delta \mathbf{r}(t)]^2\rangle \sim
  v_L (k_\perp L)^{2/3} t/L = \omega_{\mathrm{nl}} t.
\end{equation}
The equality above suggests that we could have reached the same
conclusion via the somewhat different approach of directly modifying
the time dependence of the wave modes in our turbulence: $e^{\imath
  \omega t} \rightarrow e^{\imath \omega t -
  \omega_{\mathrm{nl}} t}$ (defined for $t>0$). This
alternative approach directly describes the modes themselves as
decohering on 
a timescale $\omega_{\mathrm{nl}}^{-1}$. We will refer to this as the
\textit{exponential decorrelation} model.

To account for the uncertainty in these estimates, we replace the factor
of 1/2 in Equation \ref{eq:weinstockResonance} with $\gamma$, a
dimensionless 
order-unity constant. In this section we also choose to instead use
$\omega_{\mathrm{nl}} = ( 2\pi v_A/L ) (k_\perp L/2\pi)^{2/3}$, the
generic ``strong 
turbulence'' expression for the non-linear turnover time.\footnote{If
  one considers a simulation 
  with fixed $k$ and $v_A$ (i.e., a fixed simulation box),
  decreasing $\dot{\epsilon}$ (and therefore $v_L$) corresponds to
  weaker turbulence, since $\omega_{\mathrm{nl}} <
  \omega \sim k_\parallel v_A $.} This is for the sake of generality:
for MHD turbulence with a significant inertial range, even if the turbulence
is weak on the outer scales, the nonlinearity of the turbulence
increases on smaller scales, eventually approaching critical
balance. When critical balance is reached, the remainder of the
cascade will be in the strong regime. Furthermore, the turbulence is
weak on the outer scales, where the turbulence is being driven. Thus,
the details of the driving are more likely to be important for the
turbulence statistics in the weak regime. We present analytic
calculations of resonance broadening in weak turbulence in
Appendix \ref{app:weakStrong}.

Given Equation \ref{eq:FLRWEstimate2}, we can
perform the integral in Equation \ref{eq:weinstockResonance} to
calculate the exponential resonance function, 
\begin{equation}
  \label{eq:FLRWResonanceExp}
  R_{\mathrm{exp}}(\mathbf{k}) = \frac{\gamma v_A (k_\perp
    L)^{2/3} (2 \pi)^{1/3}}{\gamma^2 v_A^2 (2 \pi)^{2/3} (k_\perp L)^{4/3}/L +
    k_\parallel^2 L (v_\parallel \pm v_p)^2},
\end{equation}
where we have used the dispersion
relation, $\omega(\mathbf{k}) = \pm v_p |k_\parallel|$. We see that
the resonance function is still peaked at $|v_\parallel| = 
v_p$, but becomes a Lorentzian in $v_\parallel$ rather than a
delta-function.

To calculate the parallel diffusion coefficient $D_\parallel$
associated with this broadened resonance, we perform the integral in  
Equation \ref{eq:DSpectrum}, using a power spectrum of the GS form,
\begin{equation}
  \label{eq:GoldreichSridhar}
  I(\mathbf{k}) \propto v_L^2 k_\perp^{-10/3} g \left(
    \frac{k_\parallel L^{1/3}}{k_\perp^{2/3}} \right),
\end{equation}
normalized such that $v_L^2/2 = \int d^3 \mathbf{k} ~
I(\mathbf{k})/(2\pi)^3$. 
In this expression, we will treat $g(x)$ as a step function, equal to
1 if $|x| < 1$, 
and 0 otherwise, which accounts for the fact 
that power only resides in $k_\parallel \ll k_\perp$ in the inertial
range of the cascade.\footnote{The step function approximation for
  $g(x)$ 
  is for analytic convenience.   Physically, the cutoff in power in
  the $k_\parallel$ direction is unlikely to be quite so sharp.   We
  have confirmed using numerical calculations that a cutoff in
  $k_\parallel$ that is, e.g., exponential, rather than a step
  function, produces quantitatively similar results for the diffusion
  coefficients of interest.} The turbulence model described here is
broadly supported by numerical  
simulations \citep{Maron2001,Cho2002,Beresnyak2012}. However, it is
not universally accepted; see \cite{Boldyrev2006}, \cite{Perez2008},
and \cite{Grappin2010a} for discussions of 
possible shortcomings of this model. Substituting Equation
\ref{eq:GoldreichSridhar} into 
Equation \ref{eq:DSpectrum} and performing the integral leads to
\begin{equation}
  \label{eq:exponentialDecorrelation}
  \begin{split}
    D_\parallel \equiv C_1 G v_\perp^4 & \left[\frac{1}{u_-^2}\left( 1 -
        \frac{1}{u_-}
        \arctan{u_-}\right) \right.\\
    &\left.+ \frac{1}{u_+^2} \left( 1 -
        \frac{1}{u_+}
        \arctan{u_+}\right)\right],
  \end{split}
\end{equation}
where $u_\pm = (v_\parallel \pm v_p)/\gamma v_A$, $C_1$ is a
dimensionless constant, $G \equiv  v_L^2 \,
\pi \log{(L/L_{\mathrm{min}})} / (6 \, \gamma \, L \, v_A^3) $ is a
function which absorbs normalization constants,
and $L_{\mathrm{min}}$ is the smallest resolvable length scale on the
grid. We leave the normalization constant $C_1$ unspecified at the
moment; if our previous calculations were exact, then $C_1$ would be
equal to 1. We will calibrate this value against 
our test particle simulations in \S \ref{sec:particleTransport}.
Equation \ref{eq:exponentialDecorrelation} has the limiting values 
\begin{equation}
  \label{eq:limitsExp}
  D_\parallel \propto \twopartdef { \mathrm{const} } { |v_\parallel| \ll v_p} { v_p^2/v_\parallel^2} { |v_\parallel| \gg v_p.}
\end{equation}
Again, we see that the delta-function resonance predicted by linear
theory is  
substantially broadened, so that all particles with $|v_\parallel| \lesssim
v_A$ couple equally well to the turbulence; in addition,
high-velocity particles can also interact with the turbulence via the
TTD resonance.

It is not entirely clear on theoretical grounds that the
exponential decorrelation of the preceding discussion is the correct
or the only model for resonance broadening. Thus we consider also a
simple alternative, which we refer to as a \textit{Gaussian
  decorrelation} model. We replace $e^{-\gamma \omega_{\mathrm{nl}}
  t}$ in  Equation \ref{eq:GoldreichSridhar} with $e^{-(\gamma
  \omega_{\mathrm{nl}}  t)^2}$, where $\gamma$ has qualitatively the
same physical interpretation as in the exponential case. We argue that
this  functional form for the wave decoherence is more physically
motivated as it has smooth derivatives as $t \rightarrow 0$. In 
this case, we find the Gaussian resonance function, 
\begin{equation}
  \label{eq:FLRWResonanceGauss}
  \begin{split}
    R_{\mathrm{gauss}} (\mathbf{k}) & =  \frac{\pi^{1/6} L}{\gamma v_A (4 k_\perp
      L)^{2/3}} \times \\ &\exp{\left[-\frac{k_\parallel^2 L^2 (v_\parallel \pm
          v_p)^2}{4 \gamma^2 v_A^2 (2\pi)^{2/3}(k_\perp
          L)^{4/3}}\right]},
  \end{split}
\end{equation}
so that the $\delta$-function becomes a Gaussian resonance. Again we
use Equation \ref{eq:DSpectrum} to find
\begin{equation}
  \label{eq:GaussianDecorrelation}
  \begin{split}
    D_\parallel \equiv C_2 G v_\perp^4
    & \left\{ \frac{1}{u_+^3} \left[\pi \, \operatorname{erf}\left(\frac{u_-}{2}\right)
        - u_- \sqrt{\pi}  \exp{\left( - \frac{u_-^2}{4}\right)} \right] \right.\\
    + & \left. \frac{1}{u_+^3}
      \left[ \pi \, \operatorname{erf}\left(\frac{u_+}{2}\right)
        - u_+ \sqrt{\pi}  \exp{\left( - \frac{u_+^2}{4}\right)}\right] \right\}\\
  \end{split}
\end{equation}
where $\operatorname{erf}(x)$ is the error function and $C_2$ is again
a dimensionless normalization constant to be calibrated. This has the 
limiting values   
\begin{equation}
  \label{eq:limitsGauss}
  D_\parallel \propto \twopartdef { \mathrm{const} } { |v_\parallel|
    \ll v_p} { v_p^3/|v_\parallel|^3} { |v_\parallel| \gg v_p.} 
\end{equation}

\begin{figure}
  \begin{center}
    \includegraphics[width=\figwidth]{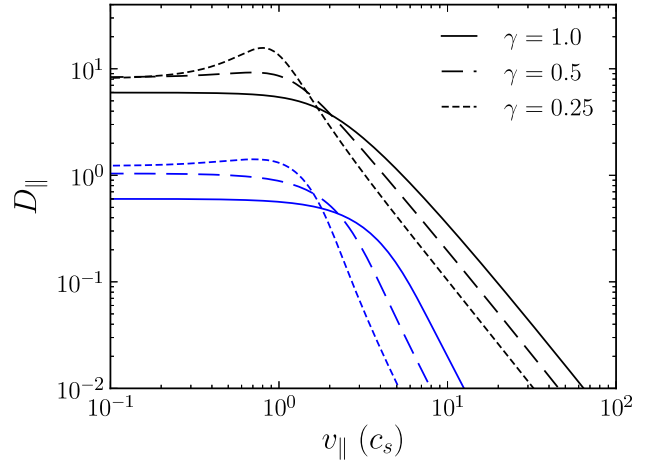}
    \caption{Analytic models of the effects of resonance broadening on
      the parallel velocity diffusion  produced by the interaction
      between particles and strong MHD  turbulence.   We show both
      exponential (black lines) and Gaussian
      (blue) models of
      resonance broadening (see \S \ref{sec:broadening}), with
      different values of $\gamma$ indicated by linestyles.  Both are
      arbitrarily normalized, with the exponential models artificially
      normalized to a higher value for clarity. $v_p = 0.81 \, c_s$ is
      chosen, consistent with $\beta=1$ slow modes. The
      dimensionless parameter 
      gamma controls the magnitude of the  broadening with gamma $\ll 1$
      approaching the linear theory  prediction of a delta function  at
      $v_\parallel = v_p$.   The exponential  resonance function gives
      $D_\parallel  
      \propto v_\parallel^{-2}$ in the high-$v_\parallel$ regime, while 
      the Gaussian model gives $D_\parallel \propto
      v_\parallel^{-3}$.}  
    \label{fig:TTD}
  \end{center}
\end{figure}

In Figure \ref{fig:TTD} we plot several representative examples of
$D_\parallel$ with arbitrary normalization. The
dimensionless parameter $\gamma$ controls the 
``peakiness'' of the resonance. Note that in all cases $D_\parallel$
declines steeply above 
$v_p$, which implies that TTD heating 
of very fast particles is inefficient. TTD acts primarily on particles
in the bulk of the plasma, near the (linear) resonance.

For completeness, we also calculate the diffusion coefficient
resulting from the linear theory delta-function resonance, with
$\omega_{\mathrm{nl}}=0$. Is this 
case, the resonance function becomes
\begin{equation}
  \label{eq:pureResonance2}
  R(\mathbf{k}) = \frac{\pi}{| k_\parallel |} \delta(v_\parallel \pm v_p).
\end{equation}
Once again, we apply the GS power spectrum to the resonant diffusion
coefficient of Equation \ref{eq:DSpectrum} to find
\begin{equation}
  \label{eq:pureResonance3}
  D_\parallel = \frac{\pi^2}{24 L} \left(\frac{v_L}{v_A}\right)^2
  v_\perp^4 \log{\left(\frac{L}{L_{\mathrm{min}}}\right)} \delta(v_\parallel \pm v_p).
\end{equation}

\subsubsection{Resonance broadening of other modes}
\label{sec:fastTTD}

\begin{figure}
  \begin{center}
    \includegraphics[width=\figwidth]{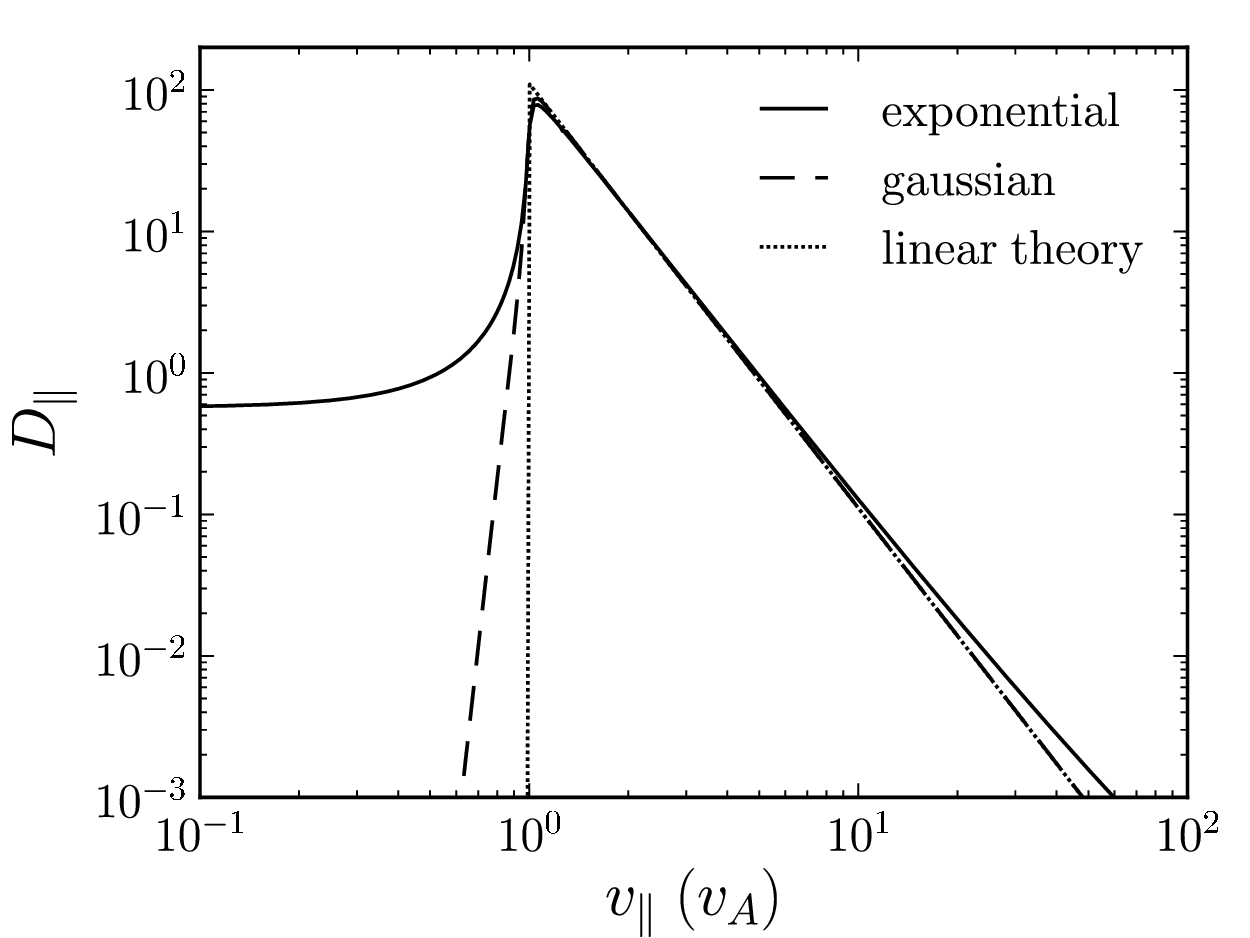}
    \caption{Fast mode contribution to resonance-broadened TTD assuming
      either an exponential or Gaussian form for the decorrelation
      (arbitrary normalization, with $\gamma = 1$; see \S
      \ref{sec:fastTTD} for more 
      details on the calculation). For comparison, we include also fast
      mode TTD    with a delta-function resonance. The weak
      non-linearity of fast modes 
      in MHD turbulence leads to a sharp resonance, very close to the
      linear theory result.}   
    \label{fig:fast}
  \end{center}
\end{figure}

Our discussion up to this point has focused on slow modes
with $\beta > 1$ for the sake of analytical simplicity. These results
are, however, relatively easy to generalize, and can be applied to
other wave modes and plasma parameter regimes. For example,
analytically accounting for fast modes is not difficult. In 
Figure \ref{fig:fast}, we plot numerically-calculated 
resonance-broadened TTD coefficients associated with fast modes in
$\beta < 1$ turbulence. We include for comparison the diffusion
coefficient predicted by a linear theory delta function
resonance. (The linear theory diffusion coefficient in this case is
not itself a delta-function due to the 
fast mode resonance condition, $k v_A = k_\parallel
v_\parallel$.) Several modifications to the derivation in  
\ref{sec:broadening} are required for the resonance-broadening
calculation. We use the fast mode dispersion 
relation, $\omega = v_A k$, and we assume an isotropic fast mode power
spectrum, $I(\mathbf{k}) \propto k^{-7/2}$ 
\citep[e.g.][]{Cho2002}. More importantly, the non-linear
decorrelation frequency
for fast modes, $v_l k$ (where $v_l$ is the turbulence velocity on
scale $l$), is much smaller than the corresponding linear
frequency $v_A k$, implying that fast modes decorrelate much more
slowly (in turbulence) than Alfv\'{e}n and slow waves. Furthermore,
this 
non-linearity becomes weaker on smaller scales. Thus, TTD with fast
modes 
will be much less broadened. 
Figure \ref{fig:fast} shows that this results in a velocity diffusion
coefficient much more peaked near $v_\parallel \sim v_A$, closer to
the linear theory result. The broadened  
resonance does still result in a power-law 
diffusion coefficient at high $v_\parallel$, with the same power-law 
indices as in the slow mode case considered above (-2 and -3 for the
exponential and Gaussian decorrelation models, respectively). However,
the distinction between the exponential and Gaussian models only
becomes apparent at very high $v_\parallel$. This
high-$v_\parallel$ tail proves to be the most important feature of
the broadened resonance for calculating heating rates at high $v_s$.

Similar modifications to the calculation in \S \ref{sec:broadening}
would need to be made in other regimes. For 
instance, for slow modes with $\beta < 1$, Equation
\ref{eq:GoldreichSridhar} would need to be modified by a
multiplicative factor $\sim \beta / (1 + \beta)$ to account for the
decreasing magnetic compression of slow modes in this
regime. Additionally, one would instead use a dispersion relation
$\omega = \pm c_s |k_\parallel|$.  However, we anticipate that the
general functional form of the diffusion coefficient is similar to
those shown in Figures \ref{fig:TTD} and \ref{fig:fast} in these
different regimes.   We will
use this generality of the resonance-broadened diffusion
coefficients to interpret our test particle simulation 
results in \S \ref{sec:particleTransport}.

\subsubsection{Type B diffusion: the $\mu \rightarrow 0$ limit}
\label{sec:FTBDiffusion}

In the $\mu \rightarrow 0$ limit, magnetic mirror forces become
negligible and all diffusion in parallel velocity is due to Fermi Type
B interactions, resulting from the tying of particles to moving,
curved field lines. In our non-relativistic case, this
change in parallel velocity in one coherent interaction with a curved
field line will typically
be of the order $\delta v_\parallel \sim v_L \sin{\theta} \sim v_L
(\delta B/B_0)$, where
$\theta$ is the opening angle of the magnetic field line, as illustrated
in Figure \ref{fig:FTB}. These 
interactions will be stochastic, and we 
estimate a parallel velocity diffusion coefficient by 
$D_\parallel \sim \delta v_{\parallel}^2/t_{\mathrm{corr}}$, where
$t_{\mathrm{corr}}$ is the typical 
time over which a particle experiences correlated field line
motion. For a particle with $| v_\parallel | \ll v_L, v_A$, the
decoherence time of interactions will be determined by the outer-scale
fluid motions. The elements of field line curvature which provide FTB
diffusion may be thought of as essentially Alfv\'{e}nic fluctuations,
because they are most effective when $\delta \mathbf{B} \perp
\mathbf{B_0}$. For strong Alfv\'{e}nic turbulence, the decoherence
time of a wave-particle interaction at the outer scale will be
provided by a combination of two effects: linear propagation and
non-linear distortion (eddy turnover). For weaker turbulence, the
linear propagation of fluctuations will control the decorrelation of
wave-particle interactions. Furthermore, at any instant in time, the
outer-scale 
fluctuations have correlation lengths $\sim \! L$, because the
turbulence is driven on this scale. Thus in either case (strong or
weak), a good estimate of the wave-particle correlation time 
is $t_{\mathrm{corr}} \sim L/v_A$, the outer-scale wave crossing
time. This allows us to estimate the diffusion coefficient
\begin{equation}
  \label{eq:lowVParFTB}
  D_\parallel \sim \frac{v_L^2}{L/v_A} \left(\frac{\delta
      B}{B_0}\right)^2 \sim \frac{v_L^4}{L \, v_A}
\end{equation}
where the second equality follows from assuming that the typical
magnetic field perturbation at the outer scale is of order $\delta
B/B_0 \sim v_L / v_A$, which will be true for Alfv\'{e}nic turbulence. 
We could choose to express this in terms of $\dot{\epsilon}$, the
turbulence cascade rate, which in Kolmogorov- or GS-like turbulence
scales as $\dot{\epsilon} \sim v_L^3/L$. However, the large-scale
eddies are those most effective at FTB acceleration. The strong
turbulence scalings are least likely to be applicable on these large
scales, and so 
we leave the FTB diffusion coefficients explicitly in terms of $v_L$.

On the other hand, particles with $|v_\parallel| \gg
v_A$ are essentially interacting with a
static snapshot of turbulence, and so $t_{\mathrm{corr}} \sim
L/|v_\parallel|$, the particle crossing time of the
outer-scale correlation length. This implies  
\begin{equation}
  \label{eq:highVParFTB}
  D_\parallel \sim \frac{v_L^2 |v_\parallel|}{L} \left(\frac{\delta
      B}{B_0}\right)^2
  \sim \frac{v_L^4}{L \, v_A^2} |v_\parallel |.
\end{equation}

We choose a functional form for $D_\parallel$ that asymptotes to the
scalings in Equations \ref{eq:lowVParFTB} and \ref{eq:highVParFTB} in
the low- and high-velocity regimes and 
varies smoothly between these limits: 
\begin{equation}
  \label{eq:FTBSimple}
  D_{\parallel,FTB} \equiv C_3 \, \frac{v_L^4}{L \, v_A} + C_4 \,
  \frac{v_L^4}{L \, v_A^2}|v_\parallel|.
\end{equation}
$C_3$ and $C_4$ are dimensionless parameters which we will
calibrate against test particle simulations in \S
\ref{sec:particleTransport}.

\subsubsection{Phase-decorrelation broadening}
\label{sec:accelerationTime}

The discussion in \S \ref{sec:broadening} focuses on the
phenomenological idea of wave-particle phase decorrelation as a result
of the decay of the resonant wave. However, the changes in parallel
velocity which the test particle experiences in the wave-particle
interaction will also lead to phase decorrelation. As a particle's
parallel velocity changes, its position $z$ along the magnetic field
changes relative to a ballistic trajectory with $z_b = z_0 +
v_{\parallel,0} \, t$. The difference $\delta z = z - z_b$ is
given by $\delta z = \int \delta v_\parallel dt$, where
$\delta v_\parallel$ is the change in particle velocity resulting from
the acceleration process. When $\delta z \sim 1/k_\parallel$, the
particle has moved completely out of phase with the wave. For
parallel velocities which change diffusively, we may
estimate that the rms change in parallel velocity is given by $\delta
v_{\parallel,\mathrm{rms}} \sim \sqrt{D_\parallel t}$. Estimating
$\delta z \sim \delta v_{\parallel,\mathrm{rms}} t$, the typical extra
random phase between wave and particle  will be $\phi \sim k_\parallel
D_\parallel^{1/2} t^{3/2} \sim  (t/t_{\mathrm{ph}})^{3/2} $, where the
final approximate equality is simply a definition of
$t_{\mathrm{ph}}$. One could now 
use this additional decorrelation in an extension of the broadened
resonance of Equation \ref{eq:weinstockResonance}. However, this
introduces a recursive dependence of $D_\parallel$ on itself, seeming
to limit the analytical tractability of this approach. For simplicity,
we do not include this effect in our analytical model. However, we
do consider this effect in interpreting our test particle results.

Using instead $\phi^2 \sim k_\parallel^2 D_\parallel
t^3$ in the resonance broadening 
calculation is formally similar to Equation 61 of 
\citet{Weinstock1969}. \citet{Yan2008a} also used a similar approach in
modelling resonance broadening. The equivalent of $t_{\mathrm{ph}}$
which they calculate may be found by using a \textit{ballistic}
approximation for the particle deviation, rather than a
\textit{diffusive} one, so that $\delta v_{\parallel,\mathrm{rms}}
\sim \mu \nabla B t$  (though this deviation is still treated as
effectively random over a distribution of particles, in that it may be
parallel or anti-parallel to the mean magnetic field), in which case
the effective $t_{\mathrm{ph}}^{-1} \sim k_\parallel v_\perp
M_A^{1/2}$. This ballistic assumption may be more be appropriate for
particle transport at early times.  (This scaling assumes that $\delta
|B| \propto M_A$, which 
is not true for pure Alfv\'{e}n waves but is the case in MHD
turbulence with a significant component of compressive energy.) This
$t_{\mathrm{ph}}$ is essentially identical to the bounce time 
for a particle of magnetic moment $\mu$ in a magnetic wave of
amplitude $M_A B_0$ and wavelength $\sim 1/k_\parallel$.

\section{Heating of a Thermal Distribution}
\label{sec:stochasticHeating}

If the evolution of a distribution of test particles
satisfies a diffusion equation (as in eqn.\ \ref{eq:fokkerPlanck}), we
may multiply both sides of this evolution equation by $m_s
v_\parallel^2/2$ and integrate over all velocities to find the
volumetric heating rate of the particles,
\begin{equation}
  \label{eq:heatingRate}
  \dot{E}_s = \frac{k_B T_s}{2 v_s^2} \int d^3v~v_\parallel^2 \frac{\partial}{\partial
    v_\parallel} \left( D_\parallel \frac{\partial f_s}{\partial v_\parallel} \right),
\end{equation}
where $T_s$ is the temperature of the particle species under
consideration, $k_B$ is Boltzmann's constant, and the distribution is
normalized such that  $\int d^3v f_s(v) = n_s$, the spatial density of
particles. For simplicity, we 
assume that the distributions are 
Maxwellian, with a thermal velocity $v_s = \sqrt{k_B T_s / m_s}$. We
will treat $T_s = T$ as a constant when we compare heating of
different particle species,
appropriate for species in temperature equilibrium.

\subsection{FTB Heating}

The parameterization of FTB diffusion in Equation \ref{eq:FTBSimple}, as
well as the assumption of a thermal distribution, 
allows us to calculate the FTB heating rate from
Equation \ref{eq:heatingRate}:
\begin{equation}
  \label{eq:FTBHeating}
  \dot{E}_{s,FTB} = \, n_{\mathrm{test}} k_B T \frac{v_L^4}{L \, v_A}\left(\frac{C_3}{v_s^2} 
    + 2 \sqrt{\frac{2}{\pi}}
    \frac{C_4}{v_s v_A} \right).
\end{equation}
$n_{\mathrm{test}}$ refers to the number density of the test particles.
The $v_s^{-2}$ dependence of this expression at small $v_s$ will cause
FTB to dominate the heating of low thermal velocity particles ($v_s
\ll c_s$, as for e.g. minor ions).

\subsection{TTD Heating: Linear Theory}

The linear theory (LT) diffusion coefficient for slow modes is given by
\begin{equation}
  \label{eq:deltaDiffusion}
  D_\parallel \propto v_\perp^4 \delta(v_\parallel \pm v_p),
\end{equation}
which gives a heating rate
\begin{equation}
  \label{eq:deltaHeating}
  \dot{E}_{s,LT} \propto T v_s^{-1} \exp{\left(-\frac{v_p^2}{2 v_s^2}\right)}.
\end{equation}
For $v_s \gg v_p$ and species at roughly the same temperature, this
heating rate scales as $v_s^{-1}$. In  the next 
section, we find that resonance broadening in general implies a
shallower dependence on $v_s$ for the asymptotic heating rate of high-$v_s$
particles. 

\subsection{TTD Heating: Resonance Broadening}

Substituting the exponential and gaussian resonance broadening
expressions into the heating rate integral in Equation \ref{eq:heatingRate}
does not lead to a 
simple analytic integral, 
and so in our comparison to our test particle simulations we will
numerically evaluate Equation \ref{eq:heatingRate}. However, as we will
see, TTD is the dominant contribution to the heating for high velocity
particles. Thus we may gain some insight by considering the heating
in the $v_\parallel \gg v_p$ limit.

In particular, the exponential decorrelation function gives
$D_\parallel \propto v_\perp^4 v_\parallel^{-2}$ for $v_\parallel \gg
v_p$. This implies a high-$v_s$ heating rate of
\begin{equation}
  \label{eq:edotExpHighV}
  \dot{E}_{s,TTD} \propto T,
\end{equation}
independent of $v_s$. Thus, the exponential model leads to more
effective heating for high-velocity particles (e.g. electrons) than linear theory. The Gaussian
decorrelation function gives 
$D_\parallel \propto v_\perp^4 |v_\parallel|^{-3}$ for $v_\parallel \gg
v_p$, on the other hand, which implies a high-$v_s$ heating rate of
\begin{equation}
  \label{eq:edotGaussHighV}
  \dot{E}_{s,TTD} \propto \frac{T \ln{(v_s)}}{v_s},
\end{equation}
which has a scaling intermediate between the LT and exponential cases,
although closer to the linear theory result given that the only
difference is the weakly varying $\ln(v_s)$ factor. 

\section{Numerical Methods}
\label{sec:numericalMethods}

Our simulations consist of collisionless test particles evolving in
isothermal, subsonic MHD turbulence. Our computational approach
is quite similar to that of
\cite{Lehe2009}, apart from two important changes noted below. We
present a summary of our methods here; more detail 
may 
be found in the earlier paper.

\subsection{The MHD Integrator}

We use the Athena MHD code of \cite{Stone2008} to evolve the
turbulence on a 3D Cartesian grid with periodic boundary
conditions. The grid is  initialized with a uniform background
magnetic field $B_0$ in the $x$-direction, with the velocity set to
zero everywhere. The initial magnitude 
of $B_0$ is set by our choice of $\beta = \rho c_s^2 / (B_0^2 / 8 \pi)
$, where $\rho$ is the fluid density and $c_s$ is the sound
speed.

We then inject kinetic energy by providing ``kicks'' to the
velocity field, in a method similar to that of
\cite{Lemaster2009}. At each timestep, we generate a velocity
perturbation $\mathbf{\delta v}(\mathbf{k})$ with random amplitudes in Fourier
space in the range of 
$2 \times \frac {2 \pi}{L} < k < 4 \times \frac {2 \pi}{L}$,
normalized by a decreasing power law in $k$, so that the majority of
the driving power enters on the largest scale of $L/2$. We also remove
modes with $|k_\parallel| < 2 \times \frac {2 \pi}{L}$, to avoid parallel
correlation lengths longer than $\sim L/2$. We enforce
$\mathbf{\delta v}(\mathbf{k}) \cdot \mathbf{k} = 0$, so that our velocity
field is divergenceless, to minimize the excitation of compressible
modes (see the discussion at the end of this subsection for more
detail on the decomposition of the turbulence in MHD modes). We
then normalize $\mathbf{\delta  v}(\mathbf{k})$ so that the net energy
input into the turbulence is given by $\dot{\epsilon}$.

We ensure that the kicks are time-correlated by
implementing an 
Ornstein-Uhlenbeck (OU) process \citep{Bartosch2001}, given by
\begin{equation}
  \label{eq:OU}
  \mathbf{\delta v} (\mathbf{k}, t + dt) = f \mathbf{\delta v}
  (\mathbf{k}, t) + \sqrt{1-f^2}\mathbf{\delta v}' (\mathbf{k}),
\end{equation}
which has an autocorrelation time (assuming the continuous limit,
$dt\rightarrow 0$) given by
\begin{equation}
  \label{eq:OUCorrelation}
  \langle \mathbf{\delta v} (\mathbf{k}, t_1) \cdot \mathbf{\delta v}
  (\mathbf{k}, t_2) \rangle = \langle [\delta v_k(t)]^2\rangle \,
  e^{(t_1-t_2)/t_{\mathrm{corr}}} 
\end{equation}
where $f = e^{-dt/t_{\mathrm{corr}}}$, $t_{\mathrm{corr}}$ is the
correlation time of the driving, $dt$ is the timestep of the driving
routine, and
$\mathbf{\delta v}'$ is a new random field generated by the process in
the previous paragraph. We
choose to drive on every MHD timestep. The OU process is simply a
mean-reverting random-walk. Note that in order for Equation \ref{eq:OUCorrelation} to
properly describe the driving statistics, the initial kick $\delta
\mathbf{v}(\mathbf{k},t=0)$ must be drawn from the same random
distribution as the subsequent $\mathbf{\delta v}' (\mathbf{k})$.

Time-correlated driving is critical for
two reasons. First, any
process that 
drives turbulence on large scales will be correlated on some typical
timescale depending on the underlying
physics of the driving process, rather than pure white
noise. Thus a time-correlated driving scheme is more representative of
the underlying physics of the turbulence. More pragmatically, evolving
our test particles in 
turbulence with $\delta(t)$-correlated driving leads to unphysical
acceleration of high gyrofrequency particles, because of the high
frequency power present in the turbulent driving.   To avoid this,
\citeauthor{Lehe2009} restricted their analysis to test particles interacting
with decaying (non-driven) turbulence. Driving via the OU process
allows us to consider particles evolving in saturated turbulence 
over arbitrary lengths of time.

On physical grounds, we choose to apply a correlation time of order 
$L/v_L$, the eddy turnover time on the outer scale of the
turbulence; see \S \ref{sec:TCorrDependence}
for a fuller investigation of the dependence of particle
heating on the correlation time. Additionally, we must
choose  $t_{\mathrm{corr}} \gg \omega_{\mathrm{max}}^{-1}$, where
$\omega_{\mathrm{max}}$ is the maximum wave mode frequency resolvable
in the MHD simulations. Smaller values of $t_{\mathrm{corr}}$ imply
essentially uncorrelated driving and lead to unphysical 
heating through a resonance with the MHD timestep. 

For a simulation with periodic boundary conditions
and a velocity field driven on the size of the domain, a particle with
arbitrarily 
high velocity effectively encounters the same eddy repeatedly, as it
crosses the box many times before the eddy decorrelates. This is
unphysical, and thus we
choose a fiducial volume for our simulations of $\{16 L, 2L, 2L\}$,
so that the box is elongated in the direction parallel to
$\mathbf{B_0}$, and there are approximately 32 uncorrelated eddies
along the length of the box. If we use instead a cubical box
of side length $2L$, we find $D_\parallel$ is unphysically affected by
box-crossing for
particles with velocities $v_\parallel \gtrsim 10 \, c_s$. Extending
the box to a size of length $16L$ in the parallel direction allows us
accurately evolve particles with velocities up to $\sim \! \! 50 \,
c_s$.  This is particularly important for studying the evolution of
electrons, corresponding to our high-velocity particles.

The parallel
extension comes at the cost of decreased resolution at the smallest
scales. However, FTB acceleration is dominated by the largest eddies,
and therefore accurately capturing smaller eddies is irrelevant to
zeroth order. Similarly, slow-mode TTD has only a logarithmic
dependence on the length of the inertial range. Thus we choose to
focus computational resources on the larger-scale eddies.

We choose $\rho = c_s = L = 1$, but the results of our simulations can
be applied to different physical systems by scaling them with
appropriate combinations of $\rho$, $c_s$, and $L$. Thus our
turbulence is controlled by three parameters: the specific energy
input rate $\dot{\epsilon}$, in units of $c_s^3 / L$; the ratio of
plasma to magnetic pressure $\beta$; and the correlation time
$t_{\mathrm{corr}}$, in units of $L / c_s$.

We note for
reference that we have applied the approximate, Fourier-space method
of \citet{Cho2003} to 
decompose the turbulent kinetic energy in our simulations into
Alfv\'{e}nic, slow, and fast mode components. Across the range of
driving rates in our simulations at fixed $\beta=1$, roughly 45\%
of the kinetic energy is in Alfv\'{e}nic modes and 45\% is in slow
modes. At lower $\beta$, an increasing fraction of the total
energy is in slow modes, up to 60\% for $\beta=0.1$, while
Alfv\'{e}nic modes lose a corresponding fraction.  Fast modes
never comprise more than $\sim$5\% percent of the 
kinetic energy, and a similarly small fraction belongs to motions with
$k_\parallel=0$ which cannot be identified with any MHD wave
mode, corresponding to interchange modes. Changing the correlation
time also results in somewhat different 2D power spectra;
specifically, longer correlation times appear to frequency-match onto
low-$k_\parallel$ modes, so that when the turbulence saturates there
is power in modes which are not directly driven. We discuss this
further in Appendix \ref{app:powerSpectra}. 

\subsection{Particle Integration}
\label{sec:particleMethods}

Our particle integration methods are described in
\cite{Lehe2009}. Once 
the turbulence reaches a fully-saturated state, particles are evolved
according to the Lorentz force. We describe particles by their
charge-to-mass ratio, expressed in the form of the mean gyrofrequency 
$\Omega_0 = q B_0 / m c$. The actual gyrofrequency of a particle will
vary according to the local value of $B$, but in subsonic turbulence,
$\delta B/ B_0 < 1$, so variations in $\Omega$ are not large. Our
simulations use ideal MHD, 
with the resistivity $\eta$ set to zero, so the turbulent dissipation is
numerical. Thus the electric field is given by $\mathbf{E} = -
\mathbf{v} \times \mathbf{B} / c$, where $\mathbf{v}$ is the fluid
velocity.

We integrate the particles with the \cite{Boris1970} implicit particle
pusher. This method is symplectic and symmetric in time, and conserves
energy and adiabatic invariants to machine precision in simulations
with constant fields in space and time. We choose a
timestep much smaller than the gyroperiod of the particle. We
interpolate the MHD fields on the grid to their value at the 
particle's 
location using the Triangular-Shaped Cloud \citep{Hockney1981} method
in space and time, 
while ensuring that the interpolation does not introduce spurious
parallel electric fields ($E_\parallel = 0$).

We initialize the particles randomly over the simulation volume,
assigning them a $v_\perp$ and $v_\parallel$, where these are measured
perpendicular and parallel to the \textit{local} magnetic
field.\footnote{Our TSC interpolation scheme means that
  the local magnetic field is measured on approximately the grid
  scale.} The perpendicular motion of a particle is the
superposition of the fast gyration around $\mathbf{B}$ and a
slowly-varying drift velocity, $v_\perp = | \mathbf{v}_{\mathrm{tot}}
- \mathbf{v}_D | $, where $\mathbf{v}_{\mathrm{tot}}$ is the total
perpendicular velocity of the particle and $\mathbf{v}_D$ is the drift
velocity. We thus require knowledge of the local drift velocity to 
accurately assign $v_\perp$. We account for the $ \mathbf{E} \times
\mathbf{B}$ drift, $\nabla B$
drift, curvature drift, and the polarization drift. These latter three
drifts are typically smaller by a factor of $\omega /
\Omega$, where $\omega^{-1}$ is the timescale of a
fluctuation, so one might naively expect them to be small in our
simulations. However, the curvature drift, approximately given as 
$v_D \sim v_\parallel^2 / \Omega R_c$ \citep{Hazeltine1998}, 
where $R_c$ is the local radius of curvature of the magnetic field
line, can become important at high $v_\parallel$.

For many of our simulations, we initialize a distribution of particles
with fixed $v_\perp$ and a logarithmically-binned distribution
in $v_\parallel$ (or vice-versa), 
to isolate the effects of one variable. In other cases, we initialize
particles according to an isotropic Maxwell-Boltzmann distribution:
\begin{equation}
  \label{eq:maxwellian}
  f (v_\parallel, v_\perp) = (2 \pi v_{s}^2)^{-3/2} \exp{\left(-
      \frac{v_\parallel^2+v_\perp^2}{2 v_s^2}\right)},
\end{equation}
where $v_s$ is the typical thermal velocity of the distribution. We
summarize fiducial parameters for our simulations in Table
\ref{table:fiducial}, and explicitly note elsewhere when different
parameters are used.

\begin{table}
  \begin{center}
    \caption{Summary of fiducial simulation parameters}
    \begin{tabular}{ c c }
      \\
      \hline \hline
      Parameter & Value \\
      \hline \hline
      $\rho$ & 1 \\
      $c_s$ & 1 \\
      $L$ & 1 \\
      Resolution & $1024\times128^2$ \\
      Volume & $16 L \times (2 L)^2$ \\
      $\dot{\epsilon}$ ($c_s^3 / L$) & 0.1\footnote{This produces a sonic Mach number of $\simeq 0.35$.} \\
      $\beta$ & 1 \\
      $N_{\mathrm{particles}}$ & $2^{11} \times 10^3 \simeq 2 \times 10^5$ \\
      $\Omega_0$ ($c_s/L$) & $10^5$ \\
      $t_{\mathrm{corr}} \, (L/c_s)$ & $1.5$
    \end{tabular}
    \label{table:fiducial}
  \end{center}
\end{table}

\section{Test Particle Diffusion in Simulations}
\label{sec:particleTransport}

We initialize a distribution of particles in $v_\parallel$, $v_\perp$
or $v$, the magnitude of the velocity. For particles initially within
a given bin in $v_\perp$ and $v_\parallel$, we calculate diffusion
coefficients according to the formal definition
\begin{equation}
  \label{eq:diffusionCoefficient}
  D_A \equiv \lim_{\delta t \to \infty} \frac{\langle (\delta
    A)^2\rangle}{2 \delta t},
\end{equation}
where $A$ is the diffusing quantity. We are typically interested in
calculating $D_\parallel$, the parallel velocity diffusion
coefficient. Our fiducial set of results are for $\beta = 1$ turbulence
with $\dot{\epsilon} = 0.1 \, c_s^3/L$ (this holds for
Figures \ref{fig:vperpDiffusion}-\ref{fig:edotsHeating}). We measure
the diffusion coefficients over a time duration from test particle
initialization until after the initial ballistic behavior has become
diffusive. For simulations presented here, this is typically
between 0.1 and 0.75 $L/c_s$ (with shorter durations for higher 
turbulent amplitudes). 

\subsection{Non-conservation of $\mu$?}
\label{sec:muNonConservation}

\begin{figure}
  \begin{center}
    \includegraphics[width=\figwidth]{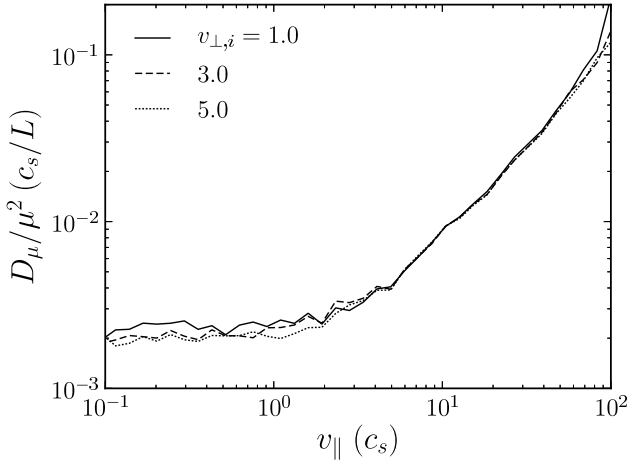}
    \caption{Diffusion coefficient for the magnetic moment $\mu$, for
      several    values of the initial 
      perpendicular velocity. Diffusion in $\mu$ is fractionally much 
      smaller than diffusion in $v_\parallel$; furthermore,
      changes in $v_\perp$ are dominated by local variations in the
      magnetic field rather than the observed slow diffusion in
      $\mu$. Thus in general we are justified in approximating the
      magnetic moment as a conserved quantity.} 
    \label{fig:vperpDiffusion}
  \end{center}
\end{figure}

We assumed throughout our analytic calculation in
\S\ref{sec:transportProperties} that $\mu$ is 
conserved. We do observe diffusive changes in $\mu$ 
throughout our simulations; i.e., $\mu$ is not in fact strictly
conserved. However, the changes in $\mu$ we find do not significantly affect our
parallel diffusion results or our interpretation of these results. In
Figure \ref{fig:vperpDiffusion}, we plot $D_\mu / \mu^2$ for our
fiducial simulation. All parallel velocities experience some 
diffusive change in $\mu$. However, this change is fractionally small 
until the highest $v_\parallel$. Furthermore, for 
particles with $v_\parallel \gg v_\perp$, velocity diffusion is primarily
due to FTB, which is independent of $\mu$, so that $\mu$ changes
significantly 
only in regimes where it is irrelevant to the dynamics. Moreover, the
diffusion coefficient for $\mu$ is fractionally much smaller than the
corresponding diffusion coefficient for $v_\parallel$. Thus we
are justified in using the approximation that $\mu$ is conserved.

\subsection{Diffusion in $v_\parallel$}
\label{sec:vParDiffusion}

\begin{figure}
  \begin{center}
    \includegraphics[width=\figwidth]{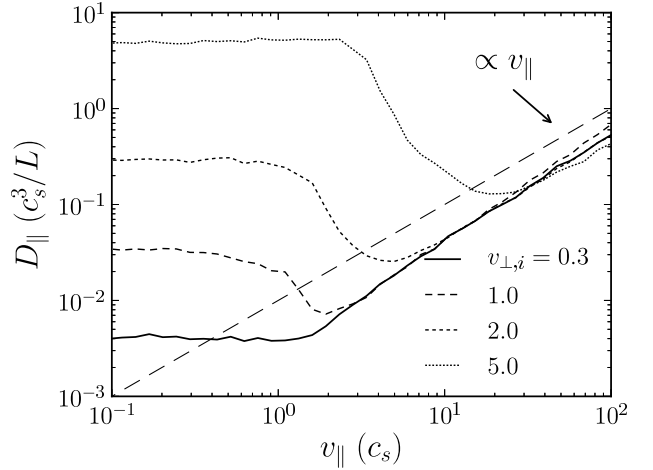}
    \caption{Parallel diffusion coefficients as a function of parallel
      velocity $v_\parallel$ for several values of
      $v_{\perp,i}$ (in units of $c_s$) for our fiducial turbulence
      properties in Table \ref{table:fiducial}. For the lowest
      $v_{\perp,i}=0.3 \, c_s$ (solid curve), 
      diffusion is provided essentially entirely by the FTB mechanism,
      which provides an effective floor for 
      $D_\parallel$ at low $v_\perp$. For higher $v_\perp$,
      resonance-broadened 
      transit-time damping
      causes further diffusion, characterized by a constant
      $D_\parallel$ below the linear theory resonance at $v_\parallel =
      v_{\mathrm{slow}} \sim c_s$, and a smooth falloff above.}
    \label{fig:DVParResonance}
  \end{center}
\end{figure}

Figure \ref{fig:DVParResonance}  shows our calculated $D_\parallel$
for particles with $v_\perp=0.3 \, c_s$ (solid line). Our fiducial
value of $\dot{\epsilon}$ is sufficiently small  that $\mu \nabla B$
forces are almost negligible for small values of $v_\perp$ and
therefore diffusion is dominated by FTB. (A similar run with
$v_\perp=0.1 \, c_s$, not shown, is essentially identical.) At low
$v_\parallel$, the diffusion coefficient saturates to a constant value
of order $v_L^4/L \,v_A \sim 0.005 \, c_s^3/L$, consistent  with
Equation \ref{eq:lowVParFTB} (though a factor of $\sim \! 4$
smaller). At  high $v_\parallel$,  the diffusion coefficient is
proportional to $v_\parallel$, consistent with our analytic derivation
in Equation \ref{eq:highVParFTB}.

Figure \ref{fig:DVParResonance} also shows that for particles with
larger $v_\perp$ (larger $\mu$),  $D_\parallel$ is significantly
larger for particles with $v_\parallel \lesssim v_{\mathrm{slow}}$,
the phase velocity of slow modes.   This  
is due to TTD, which increases in importance for larger $\mu$.   In
particular, for larger $v_\perp$, the TTD contribution manifests 
itself as an approximately constant $D_\parallel$ for $v_\parallel
\lesssim v_{\mathrm{slow}}$, 
and then as a smooth decrease in $D_\parallel$ for higher parallel
velocities. This is consistent with the resonance broadening TTD
models in \S \ref{sec:broadening}.

For particles with $v_\parallel \gg v_\perp$ beyond the linear
resonance at $v_\parallel = v_{\mathrm{slow}}$, FTB begins to again dominate the
parallel velocity diffusion.   The high $v_\parallel$ scaling of
$D_\parallel$ is $\propto v_\parallel$, consistent with Equation
\ref{eq:highVParFTB}. The importance of FTB can also be seen by the
fact that all of the curves in Fig 5 are the same at high $v_\parallel$,
independent of $v_\perp$.  This is because FTB rather than TTD provides the dominant
source of velocity diffusion at high $v_\parallel$, and the value of
$v_\perp$ ($\mu$) is irrelevant to the efficiency of FTB diffusion.

The broadened resonance in Figure \ref{fig:DVParResonance} appears to
move to the right for increasing $v_\perp$. We argue that this is the
result of phase-decorrelation broadening, discussed in
\S \ref{sec:accelerationTime}. We initialize delta-functions 
in $v_\parallel$, but as a result of the changes in $v_\parallel$
caused by finite-amplitude turbulence, these bins quickly begin to
spread out. We may set the decorrelation time of \S
\ref{sec:accelerationTime} equal to $\omega - 
k_\parallel v_\parallel = k_\parallel (v_p-v_\parallel)$ to find the
resulting broadening width $\Delta v_\parallel \equiv
v_p-v_\parallel$. The phase-decorrelation model in which the
initial particle transport is ballistic in $v_\parallel$ (as indeed we
observe at early times) predicts that $\Delta v_\parallel \sim
M_A^{1/2} v_\perp$, which is consistent with the test particle results
in Figure \ref{fig:DVParResonance}.

\begin{figure}
  \begin{center}
    \includegraphics[width=\figwidth]{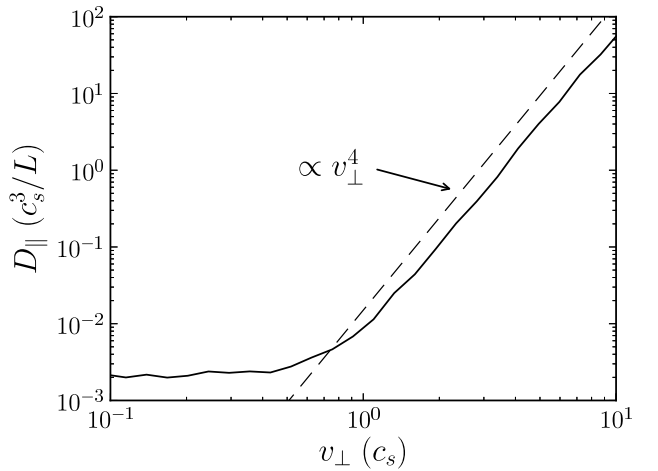}
    \caption{Parallel diffusion coefficient measured at early times as
      a      function of perpendicular velocity for a distribution of
      particles      with $v_\parallel = 0.2 \, c_s$ (other properties
      of the      simulation are summarized in Table
      \ref{table:fiducial}). At high      $v_\perp$,    $D_\parallel
      \propto v_\perp^4$; this is       due to transit-time
      damping. At low $v_\perp$, the diffusion reaches      a floor
      provided by the Fermi Type-B mechanism. At later times, we see
      shallower power laws in $v_\perp$. We believe that this is
      caused by phase-decorrelation effects; see \S
      \ref{sec:accelerationTime}.}  
    \label{fig:DScaling}
  \end{center}
\end{figure}

Figure \ref{fig:DScaling} shows $D_\parallel$ as a function
of $v_\perp$ for a distribution of particles with $v_{\parallel,i} =
0.2 \, c_s$. For high $v_\perp$, the diffusion coefficient scales like
$v_\perp^4$, consistent with the $\mu^2$ scaling in
Equation \ref{eq:exponentialDecorrelation}. This scaling may be
understood by noting that
$D_\parallel \propto a_\parallel^2 \propto (\mu \nabla B)^2$, where
$a_\parallel$ is the instantaneous acceleration felt by a charged
particle with magnetic 
moment $\mu$. For smaller $v_\perp$, as $\mu
\rightarrow 0$, the diffusion
reaches the floor provided by the FTB mechanism.  However, we 
note that at later times, we see      shallower power laws in
$v_\perp$. We believe that this is also caused by phase-decorrelation
effects due to finite changes in particle $v_\parallel$ discussed \S
\ref{sec:accelerationTime}. 

\begin{figure}
  \begin{center}
    \includegraphics[width=\figwidth]{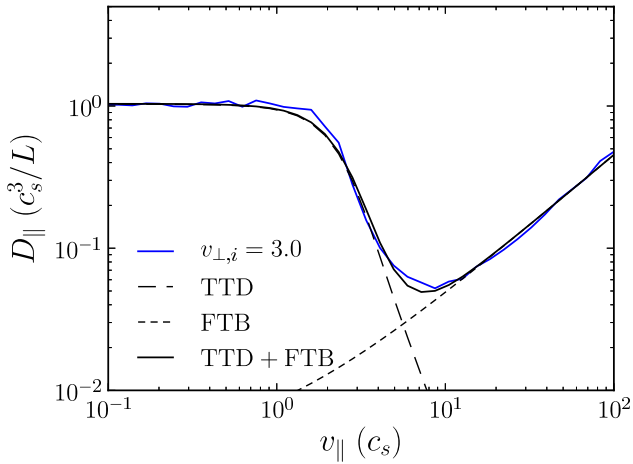}
    \caption{Comparison of numerically calculated diffusion coefficients 
      to analytical scalings (using the Gaussian decorrelation model
      with $\gamma = 0.5$),    for a simulation with 
      $v_{\perp,i}=3.0 \, c_s$. The turbulence properties are given in
      Table \ref{table:fiducial}. TTD dominates at low $v_\parallel$,
      while FTB dominates at high $v_\parallel$.}
    \label{fig:compareToTheory}
  \end{center}
\end{figure}

Figure \ref{fig:compareToTheory} presents a more quantitative
comparison between our numerical test particle results and the
analytic results for $D_\parallel$ derived in \S
\ref{sec:broadening}. The curve labeled by TTD refers to  diffusion
with a functional form provided by the Gaussian decorrelation
prescription of Equation \ref{eq:GaussianDecorrelation}. FTB  refers
to the sum of the contributions of eqs.\ \ref{eq:lowVParFTB} and
\ref{eq:highVParFTB}. The normalization of the analytic diffusion
coefficients are  chosen by eye so as to provide the best match to the
numerically determined diffusion  coefficients. Our analytic model
captures the  qualitative character of the test particle
results.

\begin{figure}
  \begin{center}
    \includegraphics[width=\figwidth]{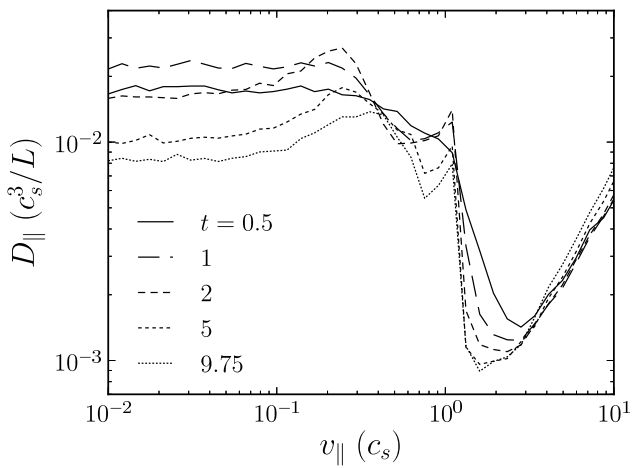}
    \caption{Parallel velocity diffusion coefficient vs. $v_\parallel$,
      measured at several different    times (in units of $L/c_s$), for
      a simulation with    $\dot{\epsilon}=0.01 c_s^3/L$, $v_{\perp,i} =
      c_s$, and $t_{\mathrm{corr}}=3 \, L/c_s$, approximately equal to
      the eddy turnover time.}  
    \label{fig:isItDiffusive1}
  \end{center}
\end{figure}

Figure \ref{fig:isItDiffusive1} shows the time dependence of our test
particle diffusion coefficients measured over different time
baselines, for a simulation with $\dot{\epsilon} = 0.01 \, c_s^3/L$
and $v_\perp = c_s$. Perfectly overlying curves measured at different
times would indicate perfectly diffusive behavior. We observe time
dependence which is generally not perfectly diffusive, and is
typically somewhat subdiffusive, in that e.g. $\langle (\delta
v_\parallel)^2 \rangle$ scales somewhat less than linearly with $t$. 

\subsection{Approach to quasilinear theory?} 
\label{sec:isItDiffusive}

For smaller turbulence amplitudes, one might in
principle expect the velocity diffusion coefficient to approach the
sharp resonance of quasilinear theory, because the turbulence becomes
increasingly weak on the outer scale. For example, in the simulation
plotted in Figure \ref{fig:isItDiffusive1}, the amplitude of the
turbulence is such that the turbulence is weak on the outer 
scale, with $v_L \simeq 0.16 \, c_s \ll v_A$. Our model of a weak
turbulence cascade in Appendix \ref{app:weakStrong} predicts a rather
sharp resonance, for these parameters. However, no obvious resonance
is present in the test particle results shown in Figure
\ref{fig:isItDiffusive1}.  

We believe that this is the result of phase-decorrelation broadening,
as discussed in \S \ref{sec:accelerationTime}. This decorrelation
effect acts in addition to the primary wave-decay decorrelation model
discussed in this paper. In turbulence, as discussed in \S
\ref{sec:accelerationTime}, there is a typical wave-particle phase
decorrelation time given by $t_{\mathrm{ph}}^{-1} \sim
k_\parallel^{2/3} D_\parallel^{1/3}$ (if the particle velocity change
is diffusive) or $t_{\mathrm{ph}}^{-1} \sim k_\parallel v_\perp
M_A^{1/2}$ (if ballistic), and we can use this to predict a broadening
width $\Delta v_\parallel$ by equating the linear frequency with the
decorrelation frequency, as in \S \ref{sec:vParDiffusion}. 

For the diffusive phase-decorrelation broadening, $\Delta v_\parallel
\sim (D_\parallel/k_\parallel)^{1/3}$.  For the results in Figure
\ref{fig:isItDiffusive1}, this approach predicts $\Delta v_\parallel
\sim 0.09 \, c_s$,
evaluated at $k_\parallel = 4 \pi/L$. This is well less than the
measured broadening. On the other hand, the ballistic
(bounce-time)  phase-decorrelation broadening gives $\Delta
v_\parallel \sim v_\perp M_A^{1/2} \sim 0.34 \, c_s$, which is consistent
with the measured broadening to within a factor of $\sim \! 2$. We
also note that the double-peaked features in $D_\parallel$ in Figure
\ref{fig:isItDiffusive1} are similar to those plotted in Appendix
\ref{app:oneWave}, where we consider the interaction of test particles
with one ideal wave. 

\subsection{Heating rate in test-particle simulations}
\label{sec:particleHeating}

In order to make a quantitative comparison between our analytical
heating rates 
and our test particle calculations, we will use the explicitly parameterized
diffusion coefficients of Equations \ref{eq:exponentialDecorrelation},
\ref{eq:GaussianDecorrelation}, and \ref{eq:FTBSimple}, corresponding
respectively to our exponential-decorrelated TTD,
Gaussian-decorrelated TTD, 
and FTB models. Throughout this section, 
we will generally  assume all species are 
at a constant reference temperature of $k T = 
\,m_s v_s^2$. Thus the important scaling is
with respect to $v_s$ (or $m_s$, equivalently). We choose $\gamma = 1$
and $\gamma = 0.5$ for the exponential and Gaussian decorrelation
models, respectively, as these values 
provide a reasonable fit to our simulations.

We simulate distributions of particles that are Maxwellian in $v =
(v_\perp^2 + v_\parallel^2)^{1/2}$, and calculate their heating rate
$\dot{E} \equiv (E_f - E_i)/(t_f -t_i)$. We normalize the
initial energy density of the distribution to $3/4 \, \rho c_s^2$, so 
that the test particles represent the energy of the ions or electrons
in a proton-electron isothermal MHD 
fluid. (Because we use test particles, all of our  heating rates may
be straightforwardly adapted to a different particle density
by multiplying by a factor $\rho_{\mathrm{test}} / \rho$.) We
calculate a numerical heating rate by fitting a straight line to the
test particle energy as a function of time, from $t=0.25-1\, L/c_s$. 

We then determine the coefficients $\{C_1, C_2,C_3,C_4\}$ in the
analytic models for $D_\parallel$ by comparison to the test particle
diffusion coefficients. For each value of $\dot{\epsilon}$, we focus
on the value of $v_\perp$ where our Gaussian analytic model best
matches the location of the high-$v_\parallel$ cutoff in the
parallel diffusion coefficient (in e.g. Figure
\ref{fig:compareToTheory}, this cutoff is around $v_\parallel \sim 2
\, c_s$). Then we choose the $C_i$ to match the   
normalization. The normalization of the analytic curve in
Figure \ref{fig:compareToTheory} is a result of this procedure.

Finally,
we calculate the associated ``analytic heating rates'' by numerically
evaluating the integral in  Equation \ref{eq:heatingRate} (integrating
the diffusion equation over parallel and perpendicular velocities)
using our fits for the dimensionless coefficients $C_i$. We 
expect the $C_i$ to depend only
weakly on parameters of the turbulence such as $\dot{\epsilon}$, $L$,
etc. In Table \ref{table:Ci}, we provide our approximate values for
these normalization coefficients for three runs at
$\dot{\epsilon}=$ 0.01, 0.1, and $1 \, c_s^3/L$, all at $\beta = 1$;
we discuss the dependence on $\beta$ in \S
\ref{sec:betaDependence}. Over a factor of 100 in $\dot{\epsilon}$, the
$C_i$ do not change significantly.

\begin{table}
  \begin{center}
    \caption{Summary of dimensionless constants $C_i$ ($\beta=1$)
      \footnote{These
        constants normalize the diffusion coefficients given by
        Equations \ref{eq:exponentialDecorrelation},
        \ref{eq:GaussianDecorrelation}, and \ref{eq:FTBSimple}.}}
    \begin{tabular}{ c c c c }
      \\
      \hline \hline
      | & $\dot{\epsilon} = 0.01$ & $\dot{\epsilon} = 0.1$ & $\dot{\epsilon} = 1$\\
      \hline
      $C_1$ & 0.085 & 0.07 & 0.065 \\
      $C_2$ & 0.15 & 0.12 & 0.11 \\
      $C_3$ & 0.25 & 0.17 & 0.2 \\
      $C_4$ & 0.5 & 0.3 & 0.4 \\

    \end{tabular}
    \label{table:Ci}
  \end{center}
\end{table}

\begin{figure}
  \begin{center}
    \includegraphics[width=\figwidth]{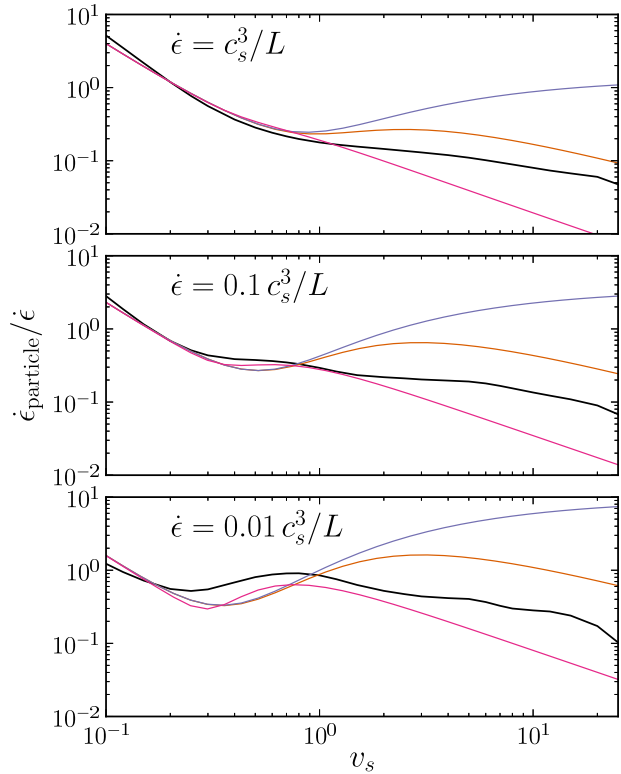}
    \caption{Test particle heating rate as a function of
      test particle thermal velocity $v_s$, for $\beta = 1$ and several
      different values     of the turbulence driving rate
      $\dot{\epsilon}$; we also show a 
      comparison to our analytically    predicted 
      heating rates from \S \ref{sec:stochasticHeating}. Black curves
      correspond to test particle results, blue curves correspond to an
      exponential decorrelation model (see eqn.\ \ref{eq:edotExpHighV}),
      orange curves correspond to
      Gaussian decorrelation (eqn.\ \ref{eq:edotGaussHighV}), and pink
      curves to a purely linear theory diffusion coefficient. In each
      case, FTB heating dominates at low $v_s$, while TTD dominates at
      high $v_s$. The Gaussian decorrelation model provides a
      notably more accurate scaling at high $v_s$ than the
      exponential model. At lower $\dot{\epsilon}$, the
      delta-function heating rate becomes more plausible; this is
      consistent with the idea that the turbulence is weaker in these
      simulations. Test    particle heating  
      of e.g. protons with $v_s = c_s/\sqrt{2}$ is within a factor of 4
      of the    turbulent cascade rate $\dot{\epsilon}$ in all cases. See
      \S \ref{sec:particleHeating} for a more detailed discussion.} 
    \label{fig:edotsHeating}
  \end{center}
\end{figure}

Figure \ref{fig:edotsHeating} shows how 
the analytic heating rates compare to our test particle results for several
different values of $\dot{\epsilon}$. Our numerically calculated
heating rates never  asymptote to a 
constant at high-$v_s$.  It is thus clear that the Gaussian
decorrelation function (orange curves) provides a better fit to
the simulation data (black curves) in the high-$v_s$ regime, 
where it scales as $\ln{(v_s)}/v_s$, as opposed to the exponential
decorrelation prescription (blue curves), which is independent
of $v_s$. Similarly, the delta function heating rate (pink curves) are
typically too steep.  In the high velocity regime, the heating rates are
reasonably well-fit by a functional form
$\dot{\epsilon}_{\mathrm{part}} \simeq  0.33 \, v_L^2
(v_s/c_s)^{-0.4}$, though this expression only applies for $\beta=1$. 

The dominant heating mechanism in Figure \ref{fig:edotsHeating} depends
on the thermal velocity of the 
test particles. Above the linear theory resonance at $v_\parallel \sim
v_{\mathrm{slow}}$, the TTD contribution to the diffusion coefficient scales like
$\mu^2 / v_\parallel^\alpha \propto v_\perp^4 / v_\parallel^\alpha$,
where $\alpha = $ 2 or 3 depending on 
the decorrelation model. Thus, in the bulk of an isotropic thermal
distribution, where $v_\perp \sim v_\parallel \sim v_s$, TTD is
increasingly important at higher $v_s$. For low thermal velocity
particles, on the other hand, $\mu \rightarrow 0$ and FTB dominates. 

As discussed in \S \ref{sec:stochasticHeating}, linear theory
implies
an asymptotic heating rate $\dot{E} \propto v_s^{-1}$ for high
$v_s$. This in turn implies that electrons, with thermal velocity $v_e
= \sqrt{m_p/m_e} \, v_p \simeq 43 \, v_p$ (where $m_p$ and $v_p$ are
respectively the mass and thermal 
velocity of protons) are heated much less effectively than protons by
the compressive fluctuations in $\beta \sim 1$ MHD turbulence. In
Table \ref{table:heatingRatio}, we provide estimates of $\dot{E}_p / 
(\dot{E}_p + \dot{E}_e)$, the proton-to-total
heating ratios in our test particle simulations, for a range of
$\dot{\epsilon}$ and $\beta$. We assign protons a value of $v_s = c_s
/\sqrt{2}$, appropriate for an equal-temperature electron-proton 
plasma. These calculations indicate that the electron heating rate is
typically smaller than the proton heating rate by a factor 
of 2-5, rather than 43. This is primarily due to the asymptotic
$\ln{(v_s)}/v_s$ scaling of the simulated 
heating rates, corresponding to our Gaussian decorrelation model (see 
eqn.\ \ref{eq:edotGaussHighV}). It is interesting to note that the
proton-to-total heating rates we find are consistent with
empirical inferences of proton vs. electron heating in the solar wind
\citep{Cranmer2009}, both in magnitude and in the increasing electron
heating for smaller $\beta$. 

\begin{table}
  \begin{center}
    \caption{Proton-to-total heating ratio in test particle simulations}
    \begin{tabular}{ c c c c }
      \\
      \hline \hline
      $\dot{E}_p / (\dot{E}_p + \dot{E}_e)$ & $\dot{\epsilon}
      (c_s^3/L)$ & $\beta$ \\ 
      \hline
      0.84 & 1 & 1 \\
      0.84 & 0.1 & 1 \\
      0.85 & 0.01 & 1 \\
      0.81 & 0.1 & 3 \\
      0.74 & 0.1 & 0.3 \\
      0.69 & 0.1 & 0.1 \\
    \end{tabular}
    \label{table:heatingRatio}
  \end{center}
\end{table}

In Figure \ref{fig:edotsHeating}, our analytic calculations with
Gaussian resonance broadening
overestimate the magnitude of the TTD heating at high $v_s$,
particularly for smaller $\dot{\epsilon}$. This is
for two reasons. The first is that we fit heating rates over the baseline
$t=[0.25,1] \, L/c_s$. However, because the distributions are driven
away from isotropy, the heating rate becomes less efficient over time,
and so the test particle energy increases somewhat sub-linearly. Our
analytics assume instead an isotropic Maxwellian at the original
thermal velocity.

A more important effect is that the analytic model used in
Figure \ref{fig:edotsHeating} is purely
strong turbulence, with $\omega_{\mathrm{nl}} \sim
\omega_{\mathrm{linear}}$ (see \S \ref{sec:broadening}). However, the
turbulence in the simulations 
with lower $\dot{\epsilon}$ is in fact weaker, with $\omega_{\mathrm{nl}} <
\omega_{\mathrm{linear}}$. We consider the diffusion coefficients
resulting from a combination of weak and strong turbulence in Appendix 
\ref{app:weakStrong}; this results in a sharper resonance which
approaches the delta-function of linear theory in the $M_A \rightarrow
0$ limit. Thus we might expect a heating rate closer
to the linear theory heating rate (the pink curves in Figure
\ref{fig:edotsHeating}) for the runs with smaller $\dot{\epsilon}$, 
as indeed is the case in Figure \ref{fig:edotsHeating}. However, this
interpretation is  
complicated by the fact that we do not find a clear peak in the
test particle velocity diffusion coefficients at low $\dot{\epsilon}$
(see Figure \ref{fig:isItDiffusive1}). This is likely due to
phase-decorrelation broadening, as discussed in \S
\ref{sec:accelerationTime}. 

\subsection{Dependence on $\beta$}
\label{sec:betaDependence}

\begin{figure}
  \begin{center}
    \includegraphics[width=\figwidth]{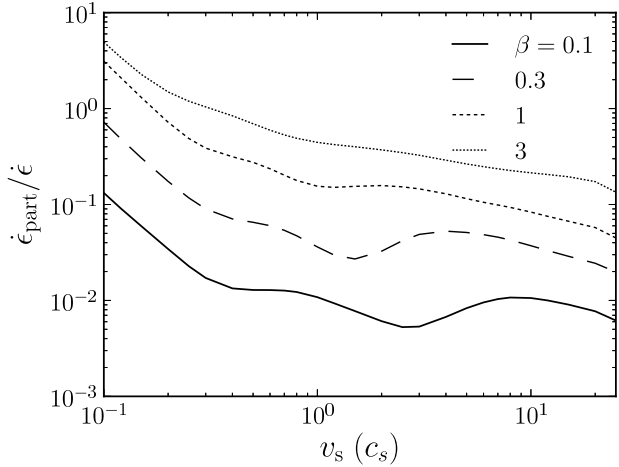}
    \caption{Test particle heating rate as a function of
      test particle thermal velocity $v_s$ for several different values
      of plasma $\beta$ at fixed kinetic energy input rate
      $\dot{\epsilon}=0.1 \, c_s^3/L$.} 
    \label{fig:betasHeating}
  \end{center}
\end{figure}

Figure \ref{fig:betasHeating} shows the simulated heating rates for a
thermal distribution of particles for
several different values of plasma $\beta$ at fixed
$\dot{\epsilon}=0.1 \, c_s^3/L$, measured over a longer baseline of
$t=[1,5] \, L/c_s$. We do not attempt to make 
a quantitative comparison with our analytic model. However,
qualitatively, three effects are clear. First, FTB heating at low
$v_s$ decreases in effectiveness at low $\beta$. This is due to the
decreasing curvature of the typical magnetic field line involved in
FTB interactions: $v_L/v_A$ decreases as $\beta$ decreases at fixed
sound speed and $\dot{\epsilon}$. 

Second, the contribution of slow modes to TTD decreases with 
decreasing beta.  The fraction of the slow mode energy in parallel
magnetic field compressions is $\propto \beta / (\beta + 1)$. Thus, at
low $\beta$ TTD heating due to the slow modes becomes less important. 

Finally, for $\beta=0.1$, there is a clear bump in the heating rate at
high thermal velocities, which moves to lower $v_s$ at higher
$\beta$. We associate this peak with TTD heating by fast modes, which
have a phase velocity approaching $v_A$ in the low-$\beta$ limit. The
reduced heating in the $\beta=0.1$ case relative to the $\beta=0.3$ is
simply due to the fact that our simulations have a greatly reduced
proportion of fast mode energy at lower $\beta$, by a factor of $\sim
\! 3$.

\subsection{Dependence on driving correlation time}
\label{sec:TCorrDependence}

\begin{figure}
  \begin{center}
    \includegraphics[width=\figwidth]{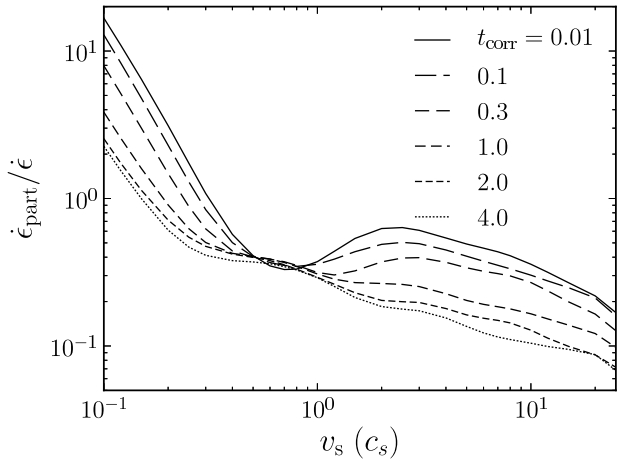}
    \caption{Test particle heating rate as a function of
      test particle thermal velocity $v_s$ for several different values
      of the OU turbulent driving 
      correlation time, $t_{\mathrm{corr}}$, in units of $L/c_s$. All other parameters are our
      fiducial parameters, listed in Table
      \ref{table:fiducial}. Changes in the heating rate are largely due
      to varying proportions of energy in the MHD modes (see \S
      \ref{sec:nonThermal}), but the overall results are consistent at
      the factor of $\sim \! 4$ level over a factor of $\sim \! 400$
      change in $t_{\mathrm{corr}}$. The large-scale eddy turnover time
      corresponds to $t_{\mathrm{corr}} \sim$ a few $L/c_s$. }
    \label{fig:TCorrHeating}
  \end{center}
\end{figure}

Figure \ref{fig:TCorrHeating} shows the heating rate of a thermal
distribution for different
$t_{\mathrm{corr}}$, the correlation time  of the Ornstein-Uhlenbeck
process with which we drive the turbulence. Over a factor of $\sim
\! 400$ in $t_{\mathrm{corr}}$, there is only a factor of $\sim \! 4$ change
in the heating rate; this suggests that to zeroth order, our results are
insensitive to the value of $t_{\mathrm{corr}}$. However, when
$t_{\mathrm{corr}} \rightarrow t_{\mathrm{MHD}}$ (not shown in Fig.\
\ref{fig:TCorrHeating}), 
where $t_{\mathrm{MHD}}$ is the MHD timestep, we find that our results
approach the limit of an uncorrelated driving scheme (typically, 
$t_{\mathrm{MHD}} \sim 8 \times 10^{-4} \, L/c_s$ in our highest resolution
simulations).  In this limit, we find acceleration
mimicking the cyclotron resonance for high-gyrofrequency particles
which should not be resonant, because of the artificially high
frequencies  introduced by the driving \citep[see
also][]{Lehe2009}.

Increasing $t_{\mathrm{corr}}$ from this minimum value while fixing
other properties of the turbulence systematically
affects the turbulent kinetic energy in each of the MHD
modes. The other physically relevant timescale in our calculations is
the outer-scale eddy  
turnover time, which for our fiducial simulation (see Table
\ref{table:fiducial}) is approximately $t_{\mathrm{eddy}} \sim L /
v_L$, which is approximately given by $\sim 3 \, L/c_s$ for the fiducial
case. Again applying the Fourier spectral decomposition of 
\cite{Cho2003},  
increasing $t_{\mathrm{corr}}$ from 0.01 $L/c_s$ to 4.0 $L/c_s$
decreases the Alfv\'{e}nic energy from 55\% to 45\%, decreases the
fast mode energy from 15\% to 3\%, and increases the slow mode energy
from 30\% to 45\% (while decreasing the overall kinetic energy in the
turbulence by roughly 30\%). Note that we hold the input power
$\dot{\epsilon}$ fixed as $t_{\mathrm{corr}}$ varies.

We also find that despite driving no
modes with parallel wavelengths longer than $L/2$, long-wavelength
modes with $k_\parallel v_A \sim t_{\mathrm{corr}}^{-1}$ naturally
appear in the developed turbulence. We believe that this is due to
frequency-matching between the correlation time of the driving and the
natural frequency of long-parallel-wavelength modes (see Appendix
\ref{app:powerSpectra}). 

These varying proportions of energy affect the thermal heating
rate in ways which are largely consistent with our interpretation
of the heating (see Fig.\ \ref{fig:TCorrHeating}). The decreasing
energy in fast modes manifests itself in a factor of $\sim \! 4$
decrease in the TTD 
heating rate at high thermal velocity, while the increase in slow mode
energy contributes to increasing the heating rate just below $v_s \sim 
c_s$.  The decrease in FTB heating is consistent with the decrease of
$v_L$ at longer correlation times.

\subsection{Non-thermal acceleration?}
\label{sec:nonThermal}

In previous sections, we have focused on short-timescale interactions
between particles and turbulence to calculate diffusion coefficients
and corresponding heating rates. On longer timescales, it is unclear
whether these interactions produce a thermal or non-thermal evolution
of the distribution function. To investigate this question, we carried
out test particle simulations lasting for longer times.

\begin{figure}
  \begin{center}
    \includegraphics[width=\figwidth]{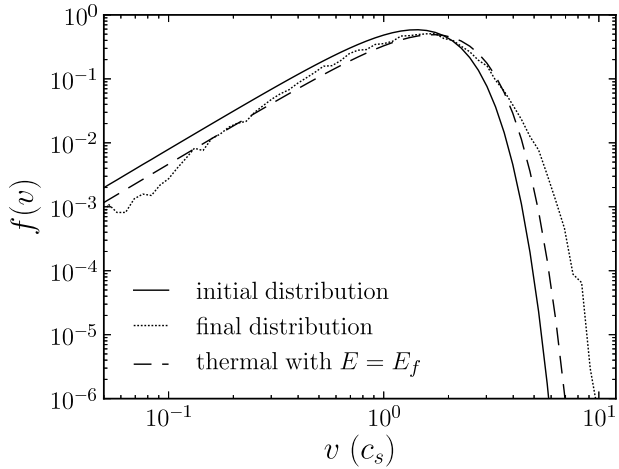}
    \caption{The distribution function of an initially thermal
      distribution of test 
      particles with $v_s= c_s$ evolving in
      turbulence with $\beta=1$ and $\dot{\epsilon}=0.1$, plotted at
      $t=0$ (solid    curve) and $t=20 \, L/c_s$ (dotted curve); this
      corresponds to $\sim \! 10$ eddy turnover times of the
      turbulence. We also plot a thermal    distribution with the same
      energy as the final distribution    (long dashed curve). The 
      distribution becomes mildly non-thermal due to the turbulent
      diffusion, but TTD  
      does not lead to a power-law tail.} 
    \label{fig:landauTail}
  \end{center}
\end{figure}

Figure \ref{fig:landauTail} shows the distribution function
(dotted curve) resulting from a simulation with $\dot{\epsilon}=0.1 \,
c_s^3/L$ run for $t=20 \, L/c_s$, 
corresponding to several eddy turnover times. The thermal energy of
the particles has
increased by $\sim \! 50 \%$. For comparison, Figure
\ref{fig:landauTail} also shows a
Maxwellian with the same energy as the final distribution (long dashed
curve). The final distribution is clearly non-thermal, in that there
is somewhat more energy in high-velocity particles, and less for
low-velocity particles. However, the distribution is only weakly
non-thermal, in that there is no evidence for the formation of a
power-law tail at high velocities. 

This weak non-thermality is not that surprising, as even our broadened
TTD resonance diffusion coefficient is largest for particles with
$v_\parallel \sim v_A$. Our result is inconsistent with recent work 
claiming that compressible turbulence generically leads to non-thermal
power law tails $f(v) \propto v^{-5}$, as observed in the solar wind
\citep[see][]{Fisk2007,Fisk2008}. However, their formalism assumes an
isotropic or nearly isotropic distribution function, enforced by
particle collisions or microscale instabilities, and has been
criticized by \cite{Jokipii2010} and \cite{Schwadron2010}, among
others. We will investigate
this point in more detail in future work.

\section{Conclusions}
\label{sec:conclusions}

We have studied the interaction between charged test particles and
low-frequency, large-scale MHD turbulence. This interaction is
important in a wide range of astrophysical systems, including the
solar wind and cosmic ray transport through the galaxy.
The coupling between particles and MHD turbulence leads to
velocity-space evolution of the particles, 
including diffusion, heating, and acceleration.
We have used simple physical arguments to
motivate analytic models of resonance broadening due to the
interaction between particles and strong turbulence (\S
\ref{sec:broadening}).   Furthermore, we 
have shown that the non-resonant 
interaction of charged particles with moving, curved magnetic field
lines (see Fig.\ \ref{fig:FTB}) is important for 
a full understanding of the velocity-space diffusion of particles in a
turbulent plasma (\S \ref{sec:FTBDiffusion}). 

We have calibrated these analytic models of
velocity space diffusion against simulations of charged test particles
in fully dynamical MHD turbulence (\S
\ref{sec:particleTransport}). These calibrations are summarized in 
Equations \ref{eq:exponentialDecorrelation} and
\ref{eq:GaussianDecorrelation}, and Table \ref{table:Ci}. We
anticipate that these calibrations of the velocity space diffusion of
test particles in MHD turbulence will be useful for a wide range of
future astrophysical and heliospheric applications.

Our most important results include:

\begin{itemize}
\item{\textit{The transit-time damping resonance is highly broadened
      in MHD turbulence,}
    relative to the delta-function prediction of linear theory. Our
    phenomenological model for resonance broadening, which 
    describes wave    decoherence in strong MHD turbulence,
    generically    leads to a velocity diffusion coefficient which
    approaches a constant at low-$v_\parallel$ and     a
    high-$v_\parallel$ power-law tail (see Fig. \ref{fig:TTD}). We
    also see evidence for phase-decorrelation broadening, in which 
    finite-amplitude turbulence accelerates particles into and out of
    resonance over relatively short timescales (see \S
    \ref{sec:accelerationTime}). The significant broadening we find
    implies     that many  
    particles, not just particles moving with the phase velocity of
    the waves, can strongly interact with the turbulence. Presumably this
    conclusion 
    would also apply to the Landau resonance in turbulence with 
    parallel electric fields, although in this study we have limited 
    our considerations and simulations to ideal    MHD.}
\item{\textit{Heating rates for high thermal velocity particles are
      inconsistent with a slow mode decorrelation model which
      is      exponential in time} (i.e., a Lorenztian resonance
    function). A Gaussian model for slow mode 
    decorrelation produces heating rates at high thermal
    velocities which are much    more consistent 
    with our test particle calculations (see Fig.\
    \ref{fig:edotsHeating}). Because of their weaker 
    nonlinearity and longer decorrelation time, we are not able to
    distinguish between the exponential and 
    Gaussian decorrelation models for fast modes.}   
\item{\textit{Fermi Type-B interactions}, wherein particles are slung
    around by moving, curved magnetic field lines (see Fig.\
    \ref{fig:FTB}), \textit{are critical      in describing the full
      velocity space diffusion of charged      particles with MHD
      turbulence} (see \S \ref{sec:FTBDiffusion} and Fig.
    \ref{fig:DVParResonance}). These interactions have a non-resonant 
    character,    and accelerate particles independent of the
    particles' magnetic    moment $\mu$. In general, FTB 
    dominates for    particles with low $\mu$ (where the TTD  
    interactions become correspondingly weak) or for particles with
    $v_\parallel \gg \left(v_A/v_L \right)^{1/2} v_\perp$
    (such that even the broadened TTD resonance has 
    fallen    off significantly).}
\item{FTB dominates the heating of particles    with thermal velocities
    $v_s$ much less than the fast and slow mode    phase speeds, 
    while TTD dominates the heating of high-$v_s$ particles (see Fig.\
    \ref{fig:edotsHeating}). FTB heating is thus particularly
    important for minor ions which have thermal velocities less than
    the plasmas sound speed. For our fiducial case, with $M_A$ similar
    to the solar wind, FTB and TTD contribute a similar amount of
    heating to protons.}  

\item{\textit{TTD can efficiently damp the turbulent
      energy in compressive MHD modes:} we find test particle heating
    rates    comparable to the turbulent energy cascade rate for a
    wide range    of plasma parameters (see Fig.\
    \ref{fig:edotsHeating}). Furthermore, electron heating    is
    comparable to proton heating for the range of $\beta
    \sim 0.1-3$ we studied (see Table
    \ref{table:heatingRatio}). Our estimated proton-to-total heating
    ratios are consistent with 
    empirical studies in the solar wind \citep{Cranmer2009}. We note,
    however, that our calculations do not include processes that damp
    the Alfvenic component of the turbulent fluctuations at small
    scales and so do not
    capture all of the heating that is likely important in the solar
    wind and other astrophysical plasmas.}  

\item{\textit{MHD turbulence does not efficiently accelerate
      collisionless test 
      particles} out of the bulk of a thermal distribution (see Fig.\
    \ref{fig:landauTail}). We find no evidence for the formation of a
    power-law tail even after many turnover times of the turbulence on
    large scales.  Instead, most of the turbulent energy is converted
    into thermal energy of the bulk of the plasma.}  
\end{itemize}

In our simulations, $\sim \! 45-50\%$ of kinetic energy is in slow
modes, while 
$\sim \! 3-5\%$ is in fast modes. As mentioned in \S
\ref{sec:numericalMethods}, this is notably higher than in the
near-Earth solar wind, where $\sim \! 10\%$ of energy is
in slow modes and a negligible fraction is in fast modes
\citep{Howes2011a}. Because compressive wave modes heat test particles
so efficiently (at a rate generally comparable to the turbulent
cascade rate $\dot{\epsilon}$), any energy initially in those modes
would be quickly damped out. This is consistent with the fact that the
solar wind contains a smaller proportion of compressive energy in the
inertial range than
naive ideal MHD simulations.

We also find that the Alfv\'{e}nic component of the turbulent cascade
will be significantly damped by Fermi Type-B interactions at the outer
scale where the turbulence spectrum is nearly isotropic. However, FTB 
interactions quickly become weak on smaller scales, and so their
effect on the the inertial range of the turbulence can probably be
neglected. Transit-time damping, on the other hand, damps energy out
of all decades in wavenumber at equal rates, but 
requires energy in compressive modes to be effective.

Another interesting consequence of FTB interactions is that velocity
diffusion depends only on the turbulence properties, and is
independent of particle mass or charge. This implies
that the heating time, $t_h \sim n k_B T / \dot{E}$, is independent of
particle mass and charge. The cooling or expansion time $t_c \sim
r/v_{\mathrm{sw}}$ in the solar wind is also independent of particle
mass and charge (where $r$ is the heliocentric distance and
$v_{\mathrm{sw}}$ is the  expansion velocity of the solar wind). If
$t_h \gg t_c$, the particles will simply cool by adiabatic expansion,
and their final temperatures will be determined by their initial
temperatures. However, if $t_h \ll t_c$, then the particles will
quickly heat up until $t_h \sim t_c$. If we assume that FTB
interactions are the most important heating process for minor ions,
this balance implies a temperature which is independent of charge but
proportional to the species mass, $T_s \sim m_s
v_L^4 r / (k_B L v_A v_{\mathrm{sw}} )$.

The interaction between particles and compressive MHD turbulence has
been invoked to explain the $v^{-5}$ distribution at high velocities
observed in the solar wind  \citep{Fisk2007,Fisk2008}. Our results are
not consistent with these models in that we see no
evidence for the development of a power-law tail to the distribution
function even after many turnover times on large scales (see Fig.\
\ref{fig:landauTail} and \S \ref{sec:nonThermal}).

In future work, we intend to implement simple pitch-angle scattering
of test particles, mimicking the effects of small-scale plasma
instabilities such as the firehose and mirror instabilities. It will be interesting to assess
whether our results on particle acceleration and the long-term
evolution of the distribution function change in the presence of
significant pitch-angle scattering. In addition, our test particle
methods are sufficiently general that they may be applied in
simulations of turbulence which are relevant on smaller scales,
e.g. Hall MHD turbulence, to probe the gyroscale transition in the
solar wind.

\begin{acknowledgements}
  We thank a very helpful anonymous referee, whose suggestions greatly
  improved the paper. This material is based on work supported by the
  National Science  Foundation Graduate Research Fellowship under
  Grant  No. DGE-1106400. Additionally, this work was supported in
  part by  NSF-DOE grant PHY-0812811, NASA HTP grant NNX11AJ37G, and
  NSF grant   ATM-0752503. Computing time 
  was provided by the National Science Foundation TeraGrid/XSEDE
  resource on the Kraken and Frost supercomputers. Additional
  computations for this paper were performed on the Henyey
  cluster at UC Berkeley, supported by NSF grant AST-090580. 
\end{acknowledgements}

\bibliographystyle{apj}
\bibliography{library,cites2}

\begin{thebibliography}{62}
\expandafter\ifx\csname natexlab\endcsname\relax\def\natexlab#1{#1}\fi

\bibitem[{Bartosch(2001)}]{Bartosch2001}
Bartosch, L. 2001, International Journal Of Modern Physics, 12, 851

\bibitem[{Beresnyak(2011)}]{Beresnyak2011c}
Beresnyak, A. 2011, Physical Review Letters, 106, 18

\bibitem[{Beresnyak(2012)}]{Beresnyak2012}
---. 2012, Monthly Notices of the Royal Astronomical Society, 422, 3495

\bibitem[{Bieber {et~al.}(1994)Bieber, Matthaeus, Smith, Wanner, Kallenrode, \&
  Wibberenz}]{Bieber1994a}
Bieber, J.~W., Matthaeus, W.~H., Smith, C.~W., Wanner, W., Kallenrode, M.-B.,
  \& Wibberenz, G. 1994, The Astrophysical Journal, 420, 294

\bibitem[{Boldyrev(2006)}]{Boldyrev2006}
Boldyrev, S. 2006, Physical Review Letters, 96, 1

\bibitem[{Boris(1970)}]{Boris1970}
Boris, J. 1970, in {Proceedings of the Fourth Conference on Numerical
  Simulations of Plasmas}, {Naval Research Lab}, 3--67

\bibitem[{Chandran(2000)}]{Chandran2000}
Chandran, B. D.~G. 2000, Physical Review Letters, 85, 4656

\bibitem[{Chandran(2003)}]{Chandran2003a}
---. 2003, The Astrophysical Journal, 599, 1426

\bibitem[{Chandran {et~al.}(2010)Chandran, Li, Rogers, Quataert, \&
  Germaschewski}]{Chandran2010}
Chandran, B. D.~G., Li, B., Rogers, B., Quataert, E., \& Germaschewski, K.
  2010, The Astrophysical Journal, 720, 503

\bibitem[{Chaston(2004)}]{Chaston2004}
Chaston, C.~C. 2004, Journal of Geophysical Research, 109, 1

\bibitem[{Chen {et~al.}(2011)Chen, Mallet, Yousef, Schekochihin, \&
  Horbury}]{Chen2011b}
Chen, C. H.~K., Mallet, A., Yousef, T.~A., Schekochihin, A.~A., \& Horbury,
  T.~S. 2011, Monthly Notices of the Royal Astronomical Society, 415, 3219

\bibitem[{Cho \& Lazarian(2002)}]{Cho2002}
Cho, J., \& Lazarian, A. 2002, Physical Review Letters, 88, 1

\bibitem[{Cho \& Lazarian(2003)}]{Cho2003}
---. 2003, Monthly Notices of the Royal Astronomical Society, 345, 325

\bibitem[{Cranmer {et~al.}(2009)Cranmer, Matthaeus, Breech, \&
  Kasper}]{Cranmer2009}
Cranmer, S.~R., Matthaeus, W.~H., Breech, B.~A., \& Kasper, J.~C. 2009, The
  Astrophysical Journal, 702, 1604

\bibitem[{Cranmer \& van Ballegooijen(2003)}]{Cranmer2003}
Cranmer, S.~R., \& van Ballegooijen, A.~A. 2003, The Astrophysical Journal,
  594, 573

\bibitem[{Cranmer \& van Ballegooijen(2012)}]{Cranmer2012}
---. 2012, The Astrophysical Journal, 754, 92

\bibitem[{Cranmer {et~al.}(2007)Cranmer, van Ballegooijen, \&
  Edgar}]{Cranmer2007}
Cranmer, S.~R., van Ballegooijen, A.~A., \& Edgar, R. 2007, The Astrophysical
  Journal Supplement Series, 171, 520

\bibitem[{Dmitruk {et~al.}(2004)Dmitruk, Matthaeus, \& Seenu}]{Dmitruk2004}
Dmitruk, P., Matthaeus, W.~H., \& Seenu, N. 2004, The Astrophysical Journal,
  617, 667

\bibitem[{Drake {et~al.}(2009)Drake, Cassak, Shay, Swisdak, \&
  Quataert}]{Drake2009}
Drake, J.~F., Cassak, P.~A., Shay, M.~A., Swisdak, M., \& Quataert, E. 2009,
  The Astrophysical Journal, 700, L16

\bibitem[{Drake {et~al.}(2006)Drake, Swisdak, Che, \& Shay}]{Drake2006}
Drake, J.~F., Swisdak, M., Che, H., \& Shay, M.~A. 2006, Nature, 443, 553

\bibitem[{Dupree(1966)}]{Dupree1966}
Dupree, T. 1966, Physics of Fluids, 9, 1773

\bibitem[{Fermi(1949)}]{Fermi1949a}
Fermi, E. 1949, Physical Review, 75, 1169

\bibitem[{Fisk \& Gloeckler(2007)}]{Fisk2007}
Fisk, L., \& Gloeckler, G. 2007, Proceedings of the National Academy of
  Sciences of the United States of America, 104, 5749

\bibitem[{Fisk \& Gloeckler(2008)}]{Fisk2008}
---. 2008, The Astrophysical Journal, 686, 1466

\bibitem[{Galtier {et~al.}(2000)Galtier, Nazarenko, Newell, \&
  Pouquet}]{Galtier2000a}
Galtier, S., Nazarenko, S.~V., Newell, A.~C., \& Pouquet, A. 2000, Journal of
  Plasma Physics, 63, 447

\bibitem[{Gary \& Borovsky(2008)}]{Gary2008}
Gary, S.~P., \& Borovsky, J.~E. 2008, Journal of Geophysical Research, 113, 1

\bibitem[{Goldreich \& Sridhar(1995)}]{Goldreich1995}
Goldreich, P., \& Sridhar, S. 1995, The Astrophysical Journal, 438, 763

\bibitem[{Goldreich \& Sridhar(1997)}]{Goldreich1997}
---. 1997, The Astrophysical Journal, 485, 680

\bibitem[{Grappin \& M\"{u}ller(2010)}]{Grappin2010a}
Grappin, R., \& M\"{u}ller, W.-C. 2010, Physical Review E, 82, 2

\bibitem[{Hazeltine \& Waelbroeck(1998)}]{Hazeltine1998}
Hazeltine, R., \& Waelbroeck, F. 1998, {The Framework of Plasma Physics}
  (Perseus Books)

\bibitem[{Higdon(1984)}]{Higdon1984}
Higdon, J.~C. 1984, The Astrophysical Journal, 285, 109

\bibitem[{Hockney \& Eastwood(1981)}]{Hockney1981}
Hockney, R., \& Eastwood, J. 1981, {Computer Simulation Using Particles} (CRC
  Press)

\bibitem[{Howes {et~al.}(2011{\natexlab{a}})Howes, Bale, Klein, Chen, Salem, \&
  TenBarge}]{Howes2011a}
Howes, G.~G., Bale, S.~D., Klein, K.~G., Chen, C. H.~K., Salem, C.~S., \&
  TenBarge, J.~M. 2011{\natexlab{a}}, 4

\bibitem[{Howes {et~al.}(2008)Howes, Cowley, Dorland, Hammett, Quataert, \&
  Schekochihin}]{Howes2008}
Howes, G.~G., Cowley, S.~C., Dorland, W., Hammett, G.~W., Quataert, E., \&
  Schekochihin, A.~A. 2008, Journal of Geophysical Research, 113, 1

\bibitem[{Howes {et~al.}(2011{\natexlab{b}})Howes, TenBarge, \&
  Dorland}]{Howes2011b}
Howes, G.~G., TenBarge, J.~M., \& Dorland, W. 2011{\natexlab{b}}, Physics of
  Plasmas, 18, 102305

\bibitem[{Jiang {et~al.}(2009)Jiang, Liu, \& Petrosian}]{Jiang2009}
Jiang, Y.~W., Liu, S., \& Petrosian, V. 2009, The Astrophysical Journal, 698,
  163

\bibitem[{Jokipii(1966)}]{Jokipii1966}
Jokipii, J.~R. 1966, The Astrophysical Journal, 146, 480

\bibitem[{Jokipii \& Lee(2010)}]{Jokipii2010}
Jokipii, J.~R., \& Lee, M.~A. 2010, The Astrophysical Journal, 713, 475

\bibitem[{Kennel \& Engelmann(1966)}]{Kennel1966}
Kennel, C., \& Engelmann, F. 1966, Physics of Fluids, 9, 2377

\bibitem[{Kraichnan(1965)}]{Kraichnan1965}
Kraichnan, R.~H. 1965, Physics of Fluids, 8, 1385

\bibitem[{Leamon {et~al.}(1999)Leamon, Smith, Ness, \& Wong}]{Leamon1999}
Leamon, R.~J., Smith, C.~W., Ness, N.~F., \& Wong, H.~K. 1999, Journal of
  Geophysical Research, 104, 22331

\bibitem[{Lehe {et~al.}(2009)Lehe, Parrish, \& Quataert}]{Lehe2009}
Lehe, R., Parrish, I.~J., \& Quataert, E. 2009, The Astrophysical Journal, 707,
  404

\bibitem[{Lemaster \& Stone(2009)}]{Lemaster2009}
Lemaster, M., \& Stone, J. 2009, The Astrophysical Journal, 691, 1092

\bibitem[{Lithwick \& Goldreich(2001)}]{Lithwick2001}
Lithwick, Y., \& Goldreich, P. 2001, The Astrophysical Journal, 562, 279

\bibitem[{Maron \& Goldreich(2001)}]{Maron2001}
Maron, J.~L., \& Goldreich, P. 2001, The Astrophysical Journal, 554, 1175

\bibitem[{McChesney {et~al.}(1987)McChesney, Stern, \& Bellan}]{McChesney1987}
McChesney, J., Stern, R., \& Bellan, P. 1987, Physical Review Letters, 59, 1436

\bibitem[{Montgomery \& Turner(1981)}]{Montgomery1981}
Montgomery, D., \& Turner, L. 1981, Physics of Fluids, 24, 825

\bibitem[{Ng \& Bhattacharjee(1997)}]{Ng1997}
Ng, C.~S., \& Bhattacharjee, A. 1997, Physics of Plasmas, 4, 605

\bibitem[{Perez \& Boldyrev(2008)}]{Perez2008}
Perez, J.~C., \& Boldyrev, S. 2008, The Astrophysical Journal, 672, L61

\bibitem[{Qin {et~al.}(2006)Qin, Matthaeus, \& Bieber}]{Qin2006}
Qin, G., Matthaeus, W.~H., \& Bieber, J.~W. 2006, The Astrophysical Journal,
  640, L103

\bibitem[{Quataert(1998)}]{Quataert1998}
Quataert, E. 1998, The Astrophysical Journal, 20, 978

\bibitem[{Quataert \& Gruzinov(1999)}]{Quataert1999}
Quataert, E., \& Gruzinov, A. 1999, The Astrophysical Journal, 520, 248

\bibitem[{Schwadron {et~al.}(2010)Schwadron, Dayeh, Desai, Fahr, Jokipii, \&
  Lee}]{Schwadron2010}
Schwadron, N.~A., Dayeh, M.~A., Desai, M., Fahr, H., Jokipii, J.~R., \& Lee,
  M.~A. 2010, The Astrophysical Journal, 713, 1386

\bibitem[{Shalchi {et~al.}(2004)Shalchi, Bieber, Matthaeus, \&
  Qin}]{Shalchi2004}
Shalchi, A., Bieber, J.~W., Matthaeus, W.~H., \& Qin, G. 2004, The
  Astrophysical Journal, 616, 617

\bibitem[{Shalchi \& Schlickeiser(2004)}]{Shalchi2004a}
Shalchi, A., \& Schlickeiser, R. 2004, Astronomy \& Astrophysics, 420, 799

\bibitem[{Shebalin {et~al.}(1983)Shebalin, Matthaeus, \&
  Montgomery}]{Shebalin1983}
Shebalin, J.~V., Matthaeus, W.~H., \& Montgomery, D. 1983, Journal of Plasma
  Physics, 29, 525

\bibitem[{Stix(1992)}]{Stix1992}
Stix, T. 1992, {Waves in Plasmas} (Springer)

\bibitem[{Stone {et~al.}(2008)Stone, Gardiner, Teuben, Hawley, \&
  Simon}]{Stone2008}
Stone, J., Gardiner, T.~A., Teuben, P., Hawley, J.~F., \& Simon, J.~B. 2008,
  The Astrophysical Journal Supplement Series, 178, 137

\bibitem[{Verdini \& Grappin(2012)}]{Verdini2012}
Verdini, A., \& Grappin, R. 2012, {Transition from weak to strong cascade in
  MHD turbulence}

\bibitem[{Weinstock(1969)}]{Weinstock1969}
Weinstock, J. 1969, Physics of Fluids, 12, 1045

\bibitem[{Yan \& Lazarian(2004)}]{Yan2004}
Yan, H., \& Lazarian, A. 2004, The Astrophysical Journal, 614, 757

\bibitem[{Yan \& Lazarian(2008)}]{Yan2008a}
---. 2008, The Astrophysical Journal, 673, 942

\end{thebibliography}

\appendix
\section{Landau-type Resonance Between Test Particles and One Wave}
\label{app:oneWave}

To consider the effects of particle trapping and bin-crossing on our
method for measuring velocity diffusion coefficients, we have
simulated a simpler self-contained one-dimensional toy problem, where
particles feel a sinusoidal, travelling acceleration of the form $a =
a_0 \sin{\left(k x - k    v_p t\right)}$. This is effectively the
Landau problem with test particles. (The toy problem is independent of
Athena, but we have confirmed that the results are qualitatively
identical to test particles interacting with a single slow mode wave
in Athena.)

\begin{figure}
  \begin{center}
    \includegraphics[width=\figwidth]{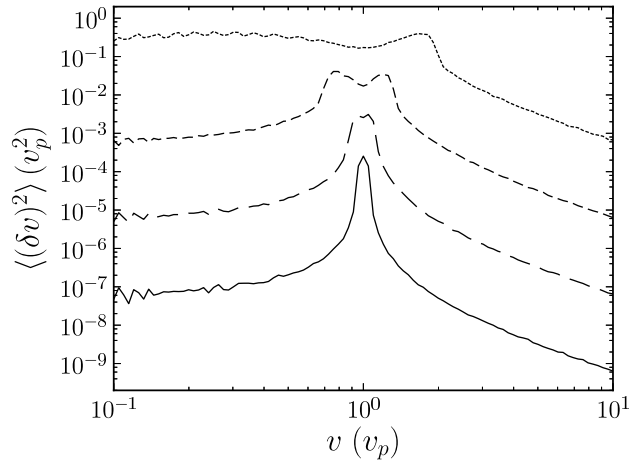}
    \caption{Mean-square velocity change for 1D toy
      problem of test particles interacting 
      with a wave of phase speed $v_p$, measured after 60 wave
      periods (long enough that the plotted quantity has
      converged). The ratio of the linear wave time to the nonlinear 
      acceleration time (evaluated for particles at the
      resonance) ranges logarithmically from 0.001 (solid curve)
      to 1 (dashed  curve). For small-amplitude waves, a sharp 
      resonance at $v_p$ is evident. As the amplitude grows larger,
      particle trapping effects associated with finite amplitude lead
      to a broadening of the resonance. A similar effect
      occurs in MHD turbulence, where finite-amplitude $\mu
      \nabla B$ forces move resonant particles off-resonance. We call
      this effect phase-decorrelation broadening (see \S
      \ref{sec:accelerationTime}).}
    \label{fig:LandauResonance}
  \end{center}
\end{figure}

The mean-square change in the velocity of test particles
$\langle (\delta v)^2 \rangle$ for several waves with different
amplitudes  is shown in Figure \ref{fig:LandauResonance}. We plot
$\langle (\delta v)^2 \rangle$ because this quantity converges to a
constant at long times. If each bin in $v$ were truly a
delta-function, and there was no numerical loss of accuracy, then
$\langle (\delta v)^2 \rangle$ for test particles in a given bin would
reach a maximum after approximately $0.5 L / (v - v_p)$ (corresponding
to half of the wave-particle 
interaction time), and then decrease back to zero as every particle
returned to its original phase with respect to the wave, though offset
by $2 \pi$. However, because particles are smoothly distributed in
velocity within each bin, slight initial velocity differences lead to
the destruction of this phase coherence over time, and $\langle
(\delta v)^2 \rangle$ converges to a constant value of roughly half
its maximum.

Acceleration by a small-amplitude wave 
produces a very sharp resonance around the wave phase velocity,
while particle trapping significantly broadens the resonance for
large-amplitude waves. For this problem, a small amplitude wave is one
for which $t_{\mathrm{lin}} / t_{\mathrm{acc}} \ll 1$, where
$t_{\mathrm{lin}} \sim 1 / v_p k$ is the linear wave period and
$t_{\mathrm{acc}} \sim v/a_0$ is the time for a particle's velocity to
change significantly due to the wave-particle interaction. The
transport in velocity space is not truly diffusive for this toy
problem, except for particles very near the resonance.

The results of this simple problem are directly analogous to test
particles interacting with MHD turbulence. When 
$t_{\mathrm{lin}} / t_{\mathrm{ph}} \gtrsim 1$ (where
$t_{\mathrm{ph}}$ is the time required for particles to significantly 
move out of phase with a wave), particles experience
significant acceleration and quickly fall out of phase with a previously
resonant wave, leading to strong resonance broadening. The situation
in turbulence is discussed in \S 
\ref{sec:accelerationTime}.

\section{Resonance Broadening in Weak Turbulence}
\label{app:weakStrong}
\subsection{Turbulence Model}
\label{app:turbulenceModel}

In the main text, we considered a resonance broadening model where the
underlying turbulence spectrum was purely strong GS
turbulence. In this appendix, we discuss an expansion of the model to
the case of weak turbulence.

We consider MHD turbulence that is driven isotropically with an outer
scale eddy velocity $v_L < v_A$.   The turbulence at the outer scale
is weak in the sense that the linear 
time scale for wave propagation is shorter than the nonlinear time
scale for eddy turnover, $L_\parallel/v_A < L_\perp/v_L$, where
isotropy implies that $L_\parallel = L_\perp \equiv L$. This implies that waves
live for more than one wave period before they decay. We may express
this weak turbulence requirement more generally by noting that the
non-linearity 
parameter $\chi \equiv v_l
l_\parallel / v_A l_\perp < 1$, where $l_\parallel$ and $l_\perp$
refer to scales smaller than $L$, and $v_l$ is the turbulent eddy
velocity on that scale.

However,
weak turbulence cascades only in the perpendicular direction to
higher $k_\perp$\citep{Shebalin1983,Galtier2000a}, while no structure on smaller
parallel scales develops. This implies a 1D power spectrum $E_k \propto
k_\perp^{-2}$\citep{Ng1997}. However, $\chi \propto k_\perp^{1/2}$, so that the
non-linearity increases at smaller scales. Eventually, when $\chi
\sim 1$, the turbulence reaches the strong critically balanced state of
GS, and the cascade proceeds to smaller scales at
fixed $\chi$ of order unity. Critical balance refers to the balance of
wave and eddy time scales. This implies that most of the turbulent
energy is contained in eddies with $k_\parallel \propto
k_\perp^{2/3}$, and a 1D power spectrum $E_k \propto
k_\perp^{-5/3}$. The transition between the weak and strong regimes
occurs at a 
wavenumber $k_{\perp,c} = (2\pi/L) M_A^{-2}$, where $M_A = v_L/v_A$ is
the Alfv\'{e}nic Mach number at the outer scale.

More quantitatively, we define the 3D power spectrum as
\begin{equation}
  \label{eq:totalPower}
  I(\mathbf{k}) =I_w(\mathbf{k}) + I_s(\mathbf{k}),
\end{equation}
where the weak and strong contributions to the power spectrum are
given respectively by 
\begin{equation}
  \label{eq:weakPower}
  I_w(\mathbf{k}) \equiv A_w k_\perp^{-3} \delta(k_\parallel - 2\pi/L)
  g\left[\frac{k_\perp}{k_{\perp,c}}\right],
\end{equation}
and 
\begin{equation}
  \label{eq:strongPower}
  I_s(\mathbf{k}) \equiv A_s k_\perp^{-10/3} g\left[\frac{k_\parallel L}{2
      \pi} \left(\frac{k_\perp}{k_{\perp,c}}\right)^{-\frac{2}{3}} \right] \left(1 - g\left[\frac{k_\perp}{k_{\perp,c}}\right]\right),
\end{equation}
where $A_w$ and $A_s$ are dimensional normalizing coefficients defined
below, and 
$g(x)$ will be treated as a step function, equal to 1 if $|x| < 1$,
and 0 otherwise. $g(x)$ is used in two ways: the first $g(x)$ in
Equation \ref{eq:strongPower} accounts for the fact that power only resides
inside the critically balanced cone of wavenumbers, while the other
instances model the transition from weak to strong turbulence at
$k_\perp = k_{\perp,c}$. 

We fix the normalizing coefficients $A_w$ and $A_s$ by first requiring
that the total 
energy in the turbulence be given by an integral over the power
spectrum,
\begin{equation}
  \label{eq:normalization}
  \frac{v_L^2}{2} = \frac{1}{(2\pi)^3} \int d^3\mathbf{k} \, I(\mathbf{k}).
\end{equation}
Because there are two
coefficients $A_w$ and $A_s$, we require another condition, which is
given by the requirement that the 1D weak and strong power spectra
must match up at the transition scale:
\begin{equation}
  \label{eq:matching}
  \left. \int d k_\parallel I_w(\mathbf{k})
  \right|_{k_{\perp,c}} 
  = \left. \int d k_\parallel I_s(\mathbf{k})
  \right|_{k_{\perp,c}}
\end{equation}
Taken together, these two requirements imply that
\begin{equation}
  \label{eq:coefficientRelation1}
  A_w = 2 A_s \left( \frac{2 \pi}{L} M_A \right)^{\frac{2}{3}}
\end{equation}
and
\begin{equation}
  \label{eq:coefficientRelation2}
  A_s = (2 \pi)^2 \frac{v_L^2}{4 F} M_A^{-\frac{2}{3}} \left(\frac{2\pi}{L}\right)^{\frac{1}{3}},
\end{equation}
where $F$ is given by 
\begin{equation}
  \label{eq:F}
  \begin{split}
    F = &
    \left(1 -
      M_A^2 \right) + \frac{3}{2}
    \left(\frac{M_A}{N}\right)^{\frac{2}{3}}  \left[
      (N M_A^2)^{\frac{2}{3}} -1 \right],
  \end{split}
\end{equation}
and $N \equiv L/L_{\mathrm{min}}$ is the range of scales in the
turbulent cascade. For $N \gg 1$, $F$ is
approximately constant and equal to 1 until $M_A$ is near unity, and
then increases sharply to $3/2$ as $M_A\rightarrow 1$. Recent  work
which demonstrates a transition from weak to strong turbulence
\citep{Howes2011b, Verdini2012} supports the basic form of this
turbulence model.  

\subsection{Resonance Broadening}
\label{app:resonanceBroadening}

We use the same resonance broadening model as in the main text (see
Equation \ref{eq:weinstockResonance}). For weak turbulence, this
approach must be modified slightly to reflect the 
fact that while $\chi$ is the fractional energy change of a wave in
one wave period, the energy change could be randomly positive or
negative, and the energy of the wave will diffuse in energy
space. Thus we estimate that the actual decoherence time of waves will
be typically given by $\omega_{\mathrm{dec}} = \chi^2
\omega_{\mathrm{lin}}$. 

As before, the exponential resonance function is given by
\begin{equation}
  \label{eq:FLRWResonanceExpApp}
  R_{\mathrm{exp}}(\mathbf{k}) = \frac{\gamma
    \omega_{\mathrm{dec}}}{\gamma^2 \omega_{\mathrm{dec}}^2 +
    k_\parallel^2 (v_\parallel \pm v_p)^2},
\end{equation}
where $\omega_{\mathrm{dec}}$ scales differently
with $k_\perp$ in the weak and strong cases. We see that
the resonance function is still peaked at $|v_\parallel| = 
v_p$, but becomes a Lorentzian in $v_\parallel$ rather than a
delta-function.

The diffusion coefficient is a sum of the contributions from the
weak and strong turbulence components. To calculate the contribution
from weak turbulence, we use the weak power spectrum of Equation
\ref{eq:weakPower}, and a non-linear frequency given by
$\omega_{\mathrm{dec}} = k_\perp v_L^2 / v_A$ to find
\begin{equation}
  \label{eq:DParWeakExp}
  \begin{split}
    D_{\parallel,w} = \frac{G'}{16 u_+^2} & \left\{
      \frac{u_+^2}{u_-^2} \ln
      \left(\frac{M_A^4+u_-^2}{M_A^4
          \left(u_-^2+1\right)}\right) \right.  \left. +\ln
      \left(\frac{M_A^4+u_+^2}{M_A^4
          \left(u_+^2+1\right)}\right) \right\}
  \end{split}
\end{equation}
where
\begin{equation}
  \label{eq:G}
  G' \equiv \frac{1}{F} \frac{2\pi}{L} \frac{v_\perp^4}{\gamma v_A}
  M_A^4,
\end{equation}
and $u_\pm = (v_\parallel \pm v_p)/(\gamma v_A)$.

\begin{figure}[h!]
  \begin{center}
    \begin{tabular}{cc}
      \includegraphics[width=\figwidth]{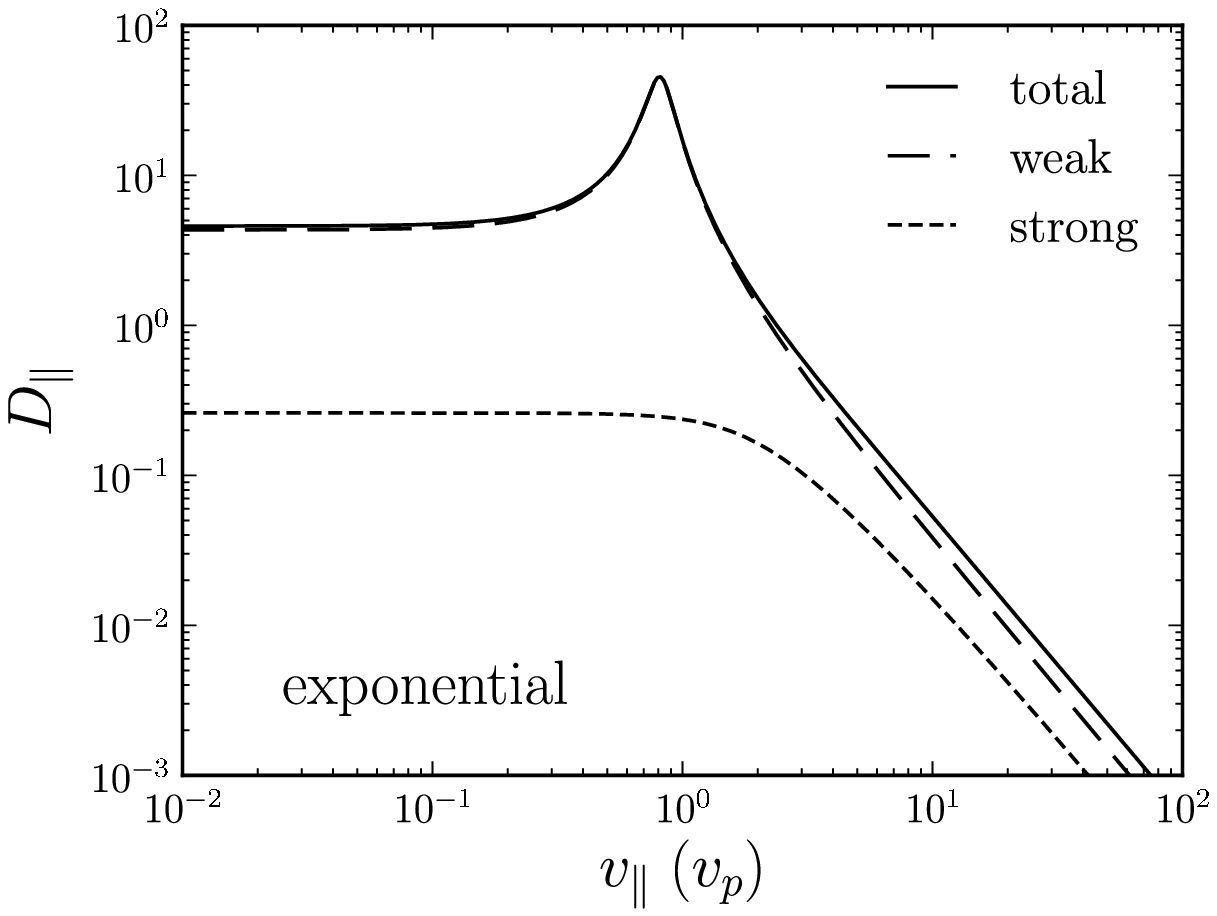}
      \includegraphics[width=\figwidth]{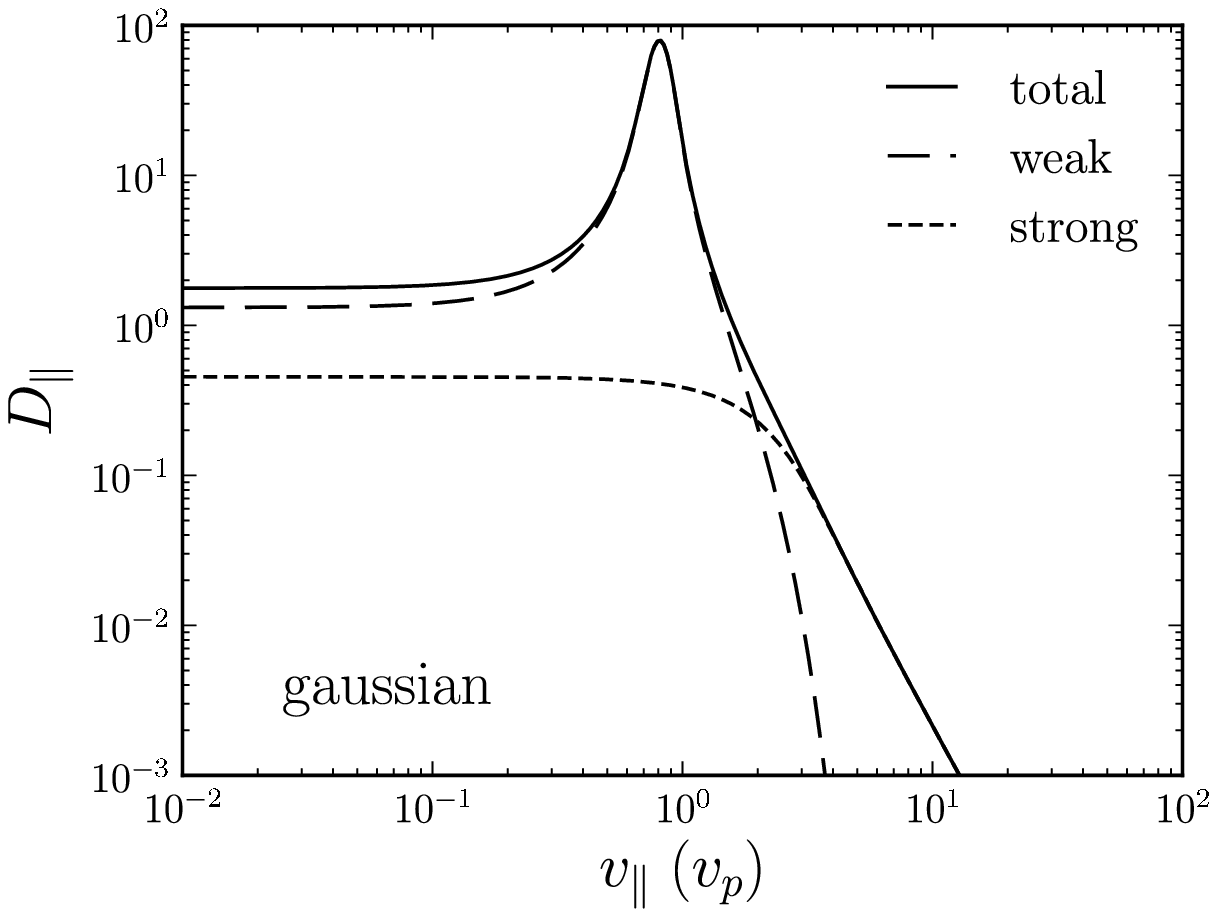}
    \end{tabular}
    \caption{Diffusion coefficients (arbitrary normalization) for
      transit-time damping by slow      modes in exponential (left)
      and Gaussian (right) decorrelation       models. $M_A = v_L/v_A
      = B_L/B_0 = 0.25$, and $v_p \simeq 0.81, v_A =      \sqrt{2}$,
      corresponding to slow modes with $\beta=1$. For this choice of
      $M_A$ we see      that weak turbulence on the outer scales is
      the dominant      contribution to the velocity diffusion,
      although the      $|v_\parallel|^{-3}$ tail from strong
      turbulence is important at      high $v_\parallel$ in the
      Gaussian model.} 
    \label{fig:weakTTD}
  \end{center}
\end{figure}

Beyond the transition to strong turbulence at $k_{\perp,c}$, the 
decorrelation frequency is given by $\omega_{\mathrm{dec}} = \omega_{\mathrm{nl}} = (2 \pi
v_A / L) (k_\perp/k_{\perp,c})^{2/3}$, and the power spectrum is given
by Equation \ref{eq:strongPower}. In this case we find
\begin{equation}
  \begin{split}
    D_{\parallel,s} & = \frac{G'}{8 u_+^2}  \ln{\left[ N M_A^2\right]}
    \left\{
      \left(1+\frac{u_+^2}{u_-^2}\right) \right.  \left. - \frac{1}{u_+}\left( \frac{u_+^3}{u_-^3}
        \arctan{u_-} + \arctan{u_+}\right) \right\},
  \end{split}
  \label{eq:DParStrongExp}
\end{equation}
which is again strongly broadened with an asymptotic form as
$|v_\parallel| \gg v_p$ of $D_{\parallel,s} \propto
v_\parallel^{-2}$. For $M_A \lesssim 0.5$, the strong turbulence
diffusion coefficient in 
Equation \ref{eq:DParStrongExp} is significantly more broadened than
the weak turbulence diffusion coefficient in Equation
\ref{eq:DParWeakExp}, as we show explicitly in Figure
\ref{fig:weakTTD} discussed below. 

As in the main text, we also examine a \textit{Gaussian
  decorrelation} model, where we replace the decorrelation term
$\gamma \omega_{\mathrm{dec}}  t$ with $(\gamma  \omega_{\mathrm{dec}}
t)^2$. In 
this case, we find the Gaussian resonance function, 
\begin{equation}
  \label{eq:FLRWResonanceGaussApp}
  R_{\mathrm{gauss}} (\mathbf{k}) = \frac{\sqrt{\pi}}{2 \gamma
    \omega_{\mathrm{dec}}} \exp{\left[-\frac{(\omega
        - k_\parallel v_\parallel)^2}{4 \gamma^2 \omega_{\mathrm{dec}}^2}\right]},
\end{equation}
so that the $\delta$-function becomes a Gaussian resonance. Now,
using Equation \ref{eq:DSpectrum}, we find 
\begin{equation}
  \label{eq:DParGaussWeak}
  \begin{split}
    D_{\parallel,w} = \frac{\sqrt{\pi }}{8} \frac{G'}{u_+^2}
    & \left\{ \frac{u_+^2}{u_-^2} \left(e^{-\frac{u_-^2}{4}}-e^{-\frac{u_-^2}{4
            M_A^4}} \right) \right.  + \left. e^{-\frac{u_+^2}{4}}-e^{-\frac{u_+^2}{
          4 M_A^4}} \right\}
  \end{split}
\end{equation}
and
\begin{equation}
  \label{eq:DParGaussStrong}
  \begin{split}
    D_{\parallel,s} \equiv & \frac{\sqrt{\pi}}{8} \frac{G'}{u_+^3} \ln{\left[N
        M_A^2\right]} \times  \left\{
      -\frac{u_+^3}{u_-^2} e^{-u_-^2/4} - u_+ e^{-u_+^2/4}  +
    \right. \left. \sqrt{\pi} \right. \left. \left[
        \frac{u_+^3}{u_-^3} \operatorname{erf} \left[u_-/2\right]  + \operatorname{erf}\left[ 
          u_+/2\right] \right] \right\}
  \end{split}
\end{equation}
where $\operatorname{erf}(x)$ is the error function. In the
$|v_\parallel| \gg v_p$ limit, $D_{\parallel,w}$ cuts off
exponentially, while $D_{\parallel,s} \propto |v_\parallel|^{-3}$.

In Figure \ref{fig:weakTTD} we plot 
representative examples of 
$D_\parallel$ for exponential and Gaussian decorrelation models, with
the same arbitrary normalization. We choose $\gamma=1$ and
$\gamma=0.5$ for the exponential and Gaussian models respectively, as
in the main text. The diffusion coefficients display a sharp peak
around the resonant phase velocity. In Figure \ref{fig:isItDiffusive1}
in the main text we plotted the corresponding diffusion coefficients
for a run with even lower amplitude turbulence ($M_A
\simeq 0.11$), for which this model would predict a yet sharper
resonance. No clear peak could be seen. We interpret that this is the
result of additional phase-decorrelation broadening due to
finite-amplitude turbulence (\S
\ref{sec:accelerationTime}), which we do not quantitatively model.

\section{Turbulence Power Spectra}
\label{app:powerSpectra}

\begin{figure}
  \begin{center}
    \includegraphics[width=6.5in]{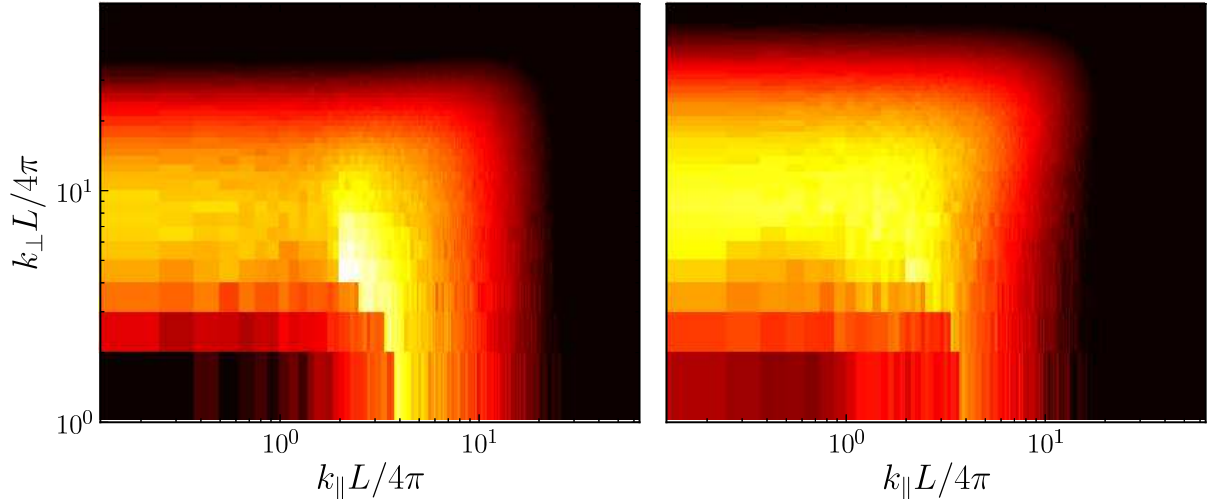}
    \caption{Two-dimensional plots of $P(\mathbf{k}) k^3$ in $\{
      k_\parallel, k_\perp\}$-space for saturated turbulence with
      $t_{\mathrm{corr}}=0.1 \, L/c_s$ (left panel) and
      $t_{\mathrm{corr}}=1.5 \, L/c_s$ (right). All other parameters
      are for our fiducial case, given in Table
      \ref{table:fiducial}.  Color normalization is
      logarithmic and identical, with white and yellow corresponding
      to the highest values, and red and black corresponding to the
      lowest values. Despite the fact that modes with
      $k_\parallel < 4\pi/L$ are not driven, frequency-matching of
      long-wavelength Alfv\'{e}n waves with the long $t_{\mathrm{corr}}$
      leads to the development of power in low-$k_\parallel$
      modes. In the run with a shorter correlation time, power
      is much more concentrated in the driven volume of $k$-space. See
      discussion in Appendix \ref{app:powerSpectra}.} 
    \label{fig:2DPowerSpectra}
  \end{center}
\end{figure}

In Figure \ref{fig:2DPowerSpectra}, we show the 2D power spectra of
fully saturated turbulence for two different simulations, one with
$t_{\mathrm{corr}}=0.1 \, L/c_s$ (left panel) and one with
$t_{\mathrm{corr}}=1.5 \, L/c_s$ right panel. In all other respects the
simulations are identical and have the fiducial parameters of Table
\ref{table:fiducial}, with an eddy turnover time of approximately
$t_{\mathrm{eddy}} \sim (L/2) / v_L \sim 1.5 \, L/c_s$.

The simulation with the longer correlation time develops significant
power on low $k_\parallel$, which are not driven by our turbulence
driving routine, while the simulation with the shorter correlation
time does not develop power on similar scales. We believe that this is
due to frequency matching of waves with the frequency content of the 
Ornstein-Uhlenbeck driving routine, so that modes with $k_\parallel
v_A \sim t_{\mathrm{corr}}^{-1}$ naturally develop power despite not
being actively driven in k-space. We also note that the
longer-$t_{\mathrm{corr}}$ simulation has a one-dimensional power
spectrum consistent with $P \propto k^{-5/3}$, while the simulation
with the shorter correlation time has a somewhat steeper power
spectrum, possibly consistent with $P \propto k^{-2}$ (though our lack
of inertial range does not allow us to make this statement with any
confidence). This seems consistent with the 
idea that the extra power at low $k_\parallel$ in the run with the longer
correlation time provides anisotropy that can actually lead to
critical balance on the outer scales. In other words, $\chi \sim
k_\perp v_L / 
k_\parallel v_A \sim 1$ because $k_\parallel/k_\perp \sim v_L/v_A$ as
a result of frequency matching. We do not
attempt to explicitly model this behavior in the main body of the
paper, however, choosing instead to focus on the simpler and more
generic case of isotropic strong turbulence. Furthermore, the test
particle diffusion coefficients do not demonstrate any obvious
peakiness for the shorter $t_{\mathrm{corr}}$ case, suggesting that
despite the apparent weaker turbulence, the diffusion coefficients are
not well-described by the turbulence model we discuss in Appendix
\ref{app:weakStrong}. We believe that this is a result of the
phase-decorrelation broadening discussed in \S
\ref{sec:accelerationTime}. 

\end{document}